\documentclass{jfm}
\usepackage{graphicx}
\usepackage{overpic}
\usepackage{epstopdf, epsfig}
\usepackage{color}
\usepackage{amsmath,bbm,bm,nicefrac}
\usepackage{pgfplots}

\newcommand\ve[1]{\boldsymbol{#1}}
\newcommand\ma[1]{\mathsfbi{#1}}

\renewcommand\ns{\hat{\bf e}_{{\rm L}1}}
\newcommand\nsc{\hat{\bf e}_{{\rm L}2}}
\newcommand\nc{\hat{\bf e}_{{\rm L}3}}
\newcommand\phim{\alpha}
\newcommand\psim{\beta}
\newcommand\ku{\text{Ku}}

\shorttitle{Alignment statistics of rods with the Lagrangian stretching direction in a channel flow}
\shortauthor{Z. Cui, A. Dubey, L. Zhao and B. Mehlig}

\title{Alignment statistics of rods with the Lagrangian stretching direction in a channel flow}
\author{Z. Cui\aff{1}\footnote[2]{These two authors contributed equally}, A. Dubey\aff{2}\footnotemark[2], L. Zhao\aff{1}\corresp{\email{zhaolihao@tsinghua.edu.cn}}  \and B. Mehlig\aff{2}}
\affiliation{\aff{1}AML, Department of Engineering Mechanics, Tsinghua University,
	100084 Beijing, China.
	\aff{2}Department of Physics, Gothenburg University,
	SE-41296 Gothenburg, Sweden}

\begin{document}
\maketitle
	
\begin{abstract}
		In homogeneous isotropic turbulence, slender rods are known to align with the Lagrangian stretching direction. However, how the degree of alignment depends on the aspect ratio of the rod is not understood.  Moreover, many flows of practical interest are anisotropic and inhomogeneous. Here we study the alignment of rods with the Lagrangian stretching direction in a channel flow, which is approximately homogeneous and isotropic near the center but inhomogeneous and anisotropic near the walls. Our main question is how the distribution of relative angles between a rod and the Lagrangian stretching direction depends on the aspect ratio of the rod and upon the distance of the rod from the channel wall. We find that the distribution exhibits two regimes: a plateau at small angles that corresponds to random uncorrelated motion, and power-law tails that describe large excursions. The variance of the relative angle is described by the width of the plateau. We find that slender rods near the channel center align better with the Lagrangian stretching direction, compared to those near the channel wall. These observations are explained in terms of simple statistical models based on Jeffery's equation, qualitatively near the channel center and quantitatively near the channel wall. Lastly we discuss the consequences of our results for the distribution of relative angles between  the orientations of nearby rods (Zhao {\em et al.}, Phys. Rev. Fluids, vol. 4, 2019, pp. 054602).
\end{abstract}

\section{Introduction}
The angular dynamics of small non-spherical particles advected in turbulence and other mixing flows is a subject of significant recent interest \citep{Wil09,Par12,Gus14,Ni14,Che13,Byron2015,Zha15,Voth16,einarsson2016a,Hejazi2017,Zha19}. In these studies it is assumed that the particles are small enough so that inertial effects  \citep{subramanian2005,einarsson2014,einarsson2015b,rosen2015d} and rotational diffusion \citep{Hinch1972} can be neglected.
In this creeping-flow limit, the equation of motion for the angular dynamics was derived by \citet{Jeffery:1922}. Jeffery's theory describes the angular dynamics of spheroidal particles in terms of their shape and the local fluid-velocity gradients.

It is usually assumed that the particles are axisymmetric, which means that they have an axis of continuous rotational symmetry, and that they possess fore-aft symmetry. In this case the main interest lies in the dynamics of the director $\ve n$ that points along the symmetry-axis of the particle. The question is how it tumbles in response to the fluid-velocity gradients. How such particles spin around their symmetry-axis $\ve n$ is usually not considered. One reason is that the spin is more difficult to measure in experiments, for axisymmetric particles. 

The shape of an axisymmetric particle with fore-aft symmetry is parameterized by its aspect ratio $\lambda=a/b$, defined here as the ratio of the symmetry-axis length $2 a$ to the diameter $2 b$ of the particle. Prolate particles have $\lambda>1$ while oblate particles have $\lambda< 1$. Jeffery considered spheroids (amongst other shapes), and showed that particle shape enters the angular dynamics in the creeping-flow limit only through the shape parameter \citep{Jeffery:1922,Bretherton:1962}:
\begin{equation}
\label{eq:SP}
\Lambda = \frac{\lambda^2-1}{\lambda^2+1}.
\end{equation}
The shape parameters $\Lambda =0, 1, -1$ correspond to spherical particles, infinitely slender rods, and infinitely thin discs, respectively.

The studies of particles in turbulence mentioned above refer to homogeneous isotropic turbulent flows, or to statistical models for such flows. One recurring observation is that the angular dynamics of axisymmetric particles  in such flows  is  quite insensitive to particle shape
for values of $|\Lambda |$ close to unity. Fig.~3c in \citet{Par12}, for instance, shows that the root mean square tumbling rate $\langle| \dot{\ve n}|^2\rangle$ in turbulence becomes
roughly  independent of shape for $|\Lambda |> 0.8$.  \citet{Ni14} show that slender rods with aspect ratio $\lambda =20$ ($\Lambda = 0.995$) follow the Lagrangian stretching direction of the turbulent flow quite closely, see Fig. 1(a) in that paper. The Lagrangian stretching direction is obtained as the leading eigenvector of the left Cauchy-Green tensor \citep{Wil09,Ni14} of the turbulent flow, which rotates exactly like a rod with infinite aspect ratio. \citet{Zha16} find that elongated rods preferentially align with the Lagrangian stretching direction, but that the orientation and  rotation behaviours appear to depend quite sensitively upon the aspect ratio near the wall \citep{Cha15,Zha15}. \citet{Par11} conclude that  in two-dimensional chaotic flows, rods preferentially align with the Lagrangian stretching direction, and that the alignment is nearly independent of the length of the rods. \citet{Deh19} consider motile bacteria with effective aspect ratio $\lambda =10$ in an inhomogeneous flow, with discrete translational symmetry, and use the Lagrangian stretching direction as a proxy for orientational alignment of the slender bacteria. \citet{Bor19} find that active particles such as motile bacteria tend to align with the instantaneous fluid velocity. Note that in this study we consider passive particles that are simply advected by the flow.
 The conclusion is, in other words, that slender rods align well with
the Lagrangian stretching direction, and that the alignment is not very sensitive to their aspect ratio in homogeneous isotropic turbulence.

One motivation for studying particles in a turbulent channel flow is that this problem is relevant for industrial applications. An example is the flow of fibre suspensions in paper making \citep{Lun11}. Such fibres are typically very slender. The fibres in the experimental study of \citet{Car10} are $0.7$\,mm long on average, with average diameter $18\,\mu$m. This corresponds to $\Lambda = 0.999$, very close to unity. Further, industrial flows are usually inhomogeneous and lack isotropy, certainly near the walls that contain them. In a turbulent channel flow near the channel center the turbulent velocity-gradient fluctuations are approximately homogeneous and isotropic, but near the wall the fluid velocity-gradient fluctuations are anisotropic in addition to a large mean shear component. We have therefore investigated the alignment statistics of relative angles between a small slender rod and the Lagrangian stretching direction in a turbulent channel flow. We study how the distribution of relative angles depends on the particle aspect ratio and on the distance of the particle from the channel wall. Our numerical studies employ direct numerical simulation (DNS) of a turbulent channel flow with a friction Reynolds number ${\rm Re}_\tau=180$ \citep{Cha15}. The diffusion approximation allows us to analytically explain the observations, qualitatively near the channel center and quantitatively near the channel wall.

Overall we find that the distribution of relative angles between the orientation of a slender rod and the Lagrangian stretching direction has a power-law tail for large angles, cut off by a plateau at small angles. When the relative angular separation between a thin rod and the Lagrangian stretching direction is small, the orientations of the two are essentially uncorrelated. However, in rare cases, the relative angle can show excursions to large angles followed by relaxation back to small angles. These large excursions gives rise to power-law tails in the distribution of the angle between the rod orientation and the Lagrangian stretching direction. The width of the plateau describes the variance of the relative angles and depends on the shape parameter, $\Lambda$, of the rod as well as the distance of the particle from the channel wall. For the same shape parameter $\Lambda$, we find that the plateau is broader near the channel wall than near the channel center, indicating that a slender rod exhibits stronger alignment with the Lagrangian stretching direction near the channel center than near the channel wall. 

We explain the observations using models for relative angles based on Jeffery's equation. The equation of motion for the relative angle is analogous to a stochastic differential equation with additive and multiplicative noises. Mathematically, the observed plateau and power-law tails are a consequence of the additive and multiplicative terms respectively. However, the physical mechanism leading to large excursions near the channel center is completely different from the mechanism near the channel wall. Near the channel center the relative angular dynamics are purely diffusive. The corresponding diffusion coefficient increases with the angular separation. Near the channel wall, on the other hand, the dynamics are a result of the weak velocity-gradient fluctuations and the strong mean shear. The weak velocity-gradient fluctuations modify both the width of the plateau and the power-law exponent in the steady state distribution of relative angles compared to the case of shear without fluctuations.

The remainder of this paper is organised as follows. In Section \ref{sec:bg} we briefly describe the background, explaining what is known about the alignment of rods in turbulent flows, introducing our notation, and describing the numerical method for direct numerical simulation (DNS) of channel flow. Section \ref{sec:results} summarizes 
the results of our DNS studies, characterizing the alignment of the particle-orientation vector $\ve n$ with the Lagrangian stretching direction near the channel center and near the wall. In Section \ref{sec:theory} we explain the observations near the channel center qualitatively and near the channel wall quantitatively by using simple models based on Jeffery's equation. 
 In Section \ref{sec:discussion} we discuss the consequences of our findings for the problem of angular structure functions. Section \ref{sec:conclusions}, finally, contains our conclusions.

\section{Background}
\label{sec:bg}
\subsection{Angular dynamics of axisymmetric particles in turbulence}
The center-of-mass of small inertialess particles simply follows the flow
\begin{align}
\label{eq:com}
 \dot{\ve x} = \ve u(\ve x(t),t)\,,
\end{align}
assuming that spatial diffusion is negligible.
Here $\ve x(t)$ is the center-of-mass position of the particle at time $t$, and $\ve u(\ve x(t),t)$ is the fluid velocity at the particle position  at time $t$. 
The orientation $\ve n$ of an axisymmetric, rigid particle with shape parameter $\Lambda$ follows Jeffery's equation \citep{Jeffery:1922}
\begin{align}
\label{eq:jeffery}
 \dot{\ve n} = \ma B(\ve x(t),t) \ve n - \big[\ve n \cdot\ma B(\ve x(t),t)\ve n\big]\ve n, 
\end{align}
with $\ma B = \ma O +\Lambda \ma S$, where $\ma O$ and $\ma S$ are the antisymmetric and symmetric parts of the fluid-gradient matrix $\ma A$.

As mentioned in the Introduction, effects of rotational diffusion and rotational inertia are neglected in Jeffery's theory.
We note that Eq.~(\ref{eq:jeffery}) holds not only for particles with continuous rotational symmetry, but also for crystals with discrete point-group symmetries \citep{Fries17,Fries18}, although there is no general formula for the parameter $\Lambda$ in terms of particle dimensions and shape.

\citet{Voth16} reviewed the angular dynamics of non-spherical particles in turbulent flows. In the following we summarize the points most relevant to our study.
Using direct numerical simulation (DNS) of a turbulent flow, \citet{Pum11} integrated Eqs.~(\ref{eq:com}) and (\ref{eq:jeffery}) for $\Lambda=1$ and concluded that 
slender rods tend to align with the vorticity vector of the turbulent flow, $\ve \omega(\ve x_t,t) = \ve\nabla \times \ve u(\ve x_t,t)$. They explained this qualitatively by noting that
the equation of motion for $\ve\omega$ resembles Eq.~(\ref{eq:jeffery}). Many authors \citep{Guala,Par12,Pum11,Gus14,Ni14,Che13} have analysed how $\ve\omega$ and $\ve n$ align with the orthogonal system
of eigenvectors of the local strain-rate matrix $\ma S(\ve x_t,t)$. These DNS studies show that $\ve n$ and also $\ve\omega$ tend to align to some extent with the middle eigenvector of the strain-rate matrix.
This is surprising, because it is natural to expect that these vectors might align with the maximal eigendirection of $\ma S(\ve x_t,t)$. This puzzle was resolved
by  \citet{xu2011} who explained that $\ve n(t)$ tends to follow the maximal eigendirection of the strain-rate matrix, but that the eigensystem of $\ma S(\ve x_t,t)$ rotates away as $\ve n(t)$ turns. In other words, the complex angular
dynamics w.r.t. the eigensystem of the strain-rate matrix is a consequence of the fact that the time scales of the turbulent dynamics and that of $\ve n(t)$ are similar. 
In summary, the particle-orientation vector of a slender rod in turbulence preferentially samples certain directions defined by the local turbulent velocity gradients, but the picture is quite intricate, even for a single slender rod.
\citet{Byron2015} analysed the angular dynamics of slender disks in turbulent flow, and found that the symmetry vector tends to stay in the plane orthogonal to vorticity. This has important consequences for how slender rods and thin disks
tumble in turbulent flows \citep{Par12,Gus14}.

A simpler picture emerges if one describes local alignment of the rod direction with the eigensystem of the left Cauchy-Green tensor \citep{Wil09,Wil10a,Wil11,Ni14,Hejazi2017}, simply because the direction of infinitely slender rods must converge to the leading eigenvector of this tensor. In short, one defines the deformation tensor $\ma D$ as the solution of the differential equation
\begin{align}
 \frac{{\rm d}}{{\rm d}t} \ma D(t)&= \ma A(t) \ma D(t)\,, \quad\mbox{with initial condition}\quad
 \ma D(0) = \mathbbm{1}\,.
\end{align}
The left Cauchy-Green tensor $\ma M(t)$ is then formed  as 
$\ma M = \ma D \ma D^{\sf T} $. The tensor $\ma M(t)$ is  symmetric with eigenvalues $\sigma_1(t) > \sigma_2(t)> \sigma_3(t)> 0$, and eigenvectors $\ns(t), \nsc(t), \nc(t)$. The leading eigenvector $\ns(t)$ corresponding to the largest eigenvalue $\sigma_1$ is called {\em  Lagrangian stretching direction}, while the eigenvector $\nc$ corresponding to the smallest eigenvalue $\sigma_3$ is the compressing direction. The middle eigenvector, $\nsc$, forms an orthonormal coordinate system with the other two.

In the long time limit, the equation of motion of the Lagrangian stretching direction $\ns(t)$ \citep{Bal99} reduces to the equation for $\ve n(t)$ given by Eq.~(\ref{eq:jeffery}) for an infinitely slender rod, $\Lambda=1$. Further it can be argued that in the steady state the difference in the two orientations must be zero, and we have checked this in simulations.

\begin{figure}
\centering
\vspace*{2mm}
         \begin{overpic}[scale=0.4]{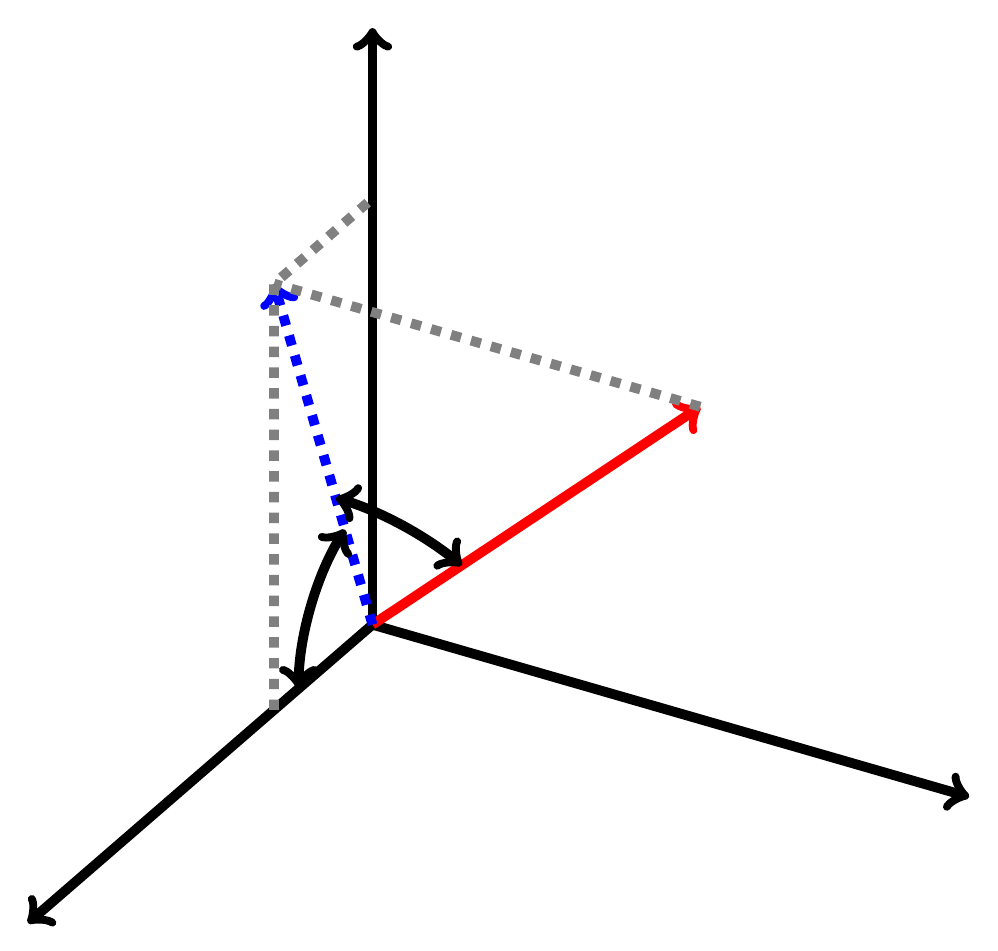}
         \put(-5,84.9){\colorbox{white}{(a)}}
		\put(72,55){$\ve n$}
		 \put(100,13){$\nsc$}
                \put(-4,0){$\ns$}
                \put(40,90){$ \nc$}
		\put(27.7,39.8){{$\alpha$}}
		\put(42,43.7){$\beta$}
	\end{overpic}\hspace*{1cm}
         \begin{overpic}[scale=0.4]{./figures/coordinates_wall}
         \put(-5,84.9){\colorbox{white}{(b)}}
		\put(72,55){$\ve n$}
		\put(100,13){$\hat{\bf{y}}$}
		\put(-4,0){$\hat{\bf{x}}$}
		\put(40,90){$\hat{\bf{z}}$}
		\put(27.7,39.8){{$\phi$}}
		\put(42,43.7){$\theta$}
	\end{overpic}
     \mbox{}\vspace*{3mm}
 \caption{Euler angles  used in analyzing the alignment of the particle-symmetry vector $\ve n$. (a) Coordinate system defined by the eigenvectors   $\ns(t), \nsc(t), \nc(t)$ of the left Cauchy-Green tensor $\ma M(t)$. 
  Here $\ns$ and $\nc$ are the expanding and contracting directions, while
 $\nsc$ is chosen to keep the coordinate system right-handed, and $\phim$, $\psim$ are Euler angles in this reference frame. 
(b) Fixed Cartesian channel-coordinate system with basis vectors  $\hat{\bf x}, \hat{\bf y}$, and $\hat{\bf z}$.
Here $\hat{\bf x}$ is the  streamwise direction, $\hat{\bf y}$ the span-wise, and $\hat{\bf z}$ the wall-normal direction of the channel flow. The Euler angles are
$\phi$ and $\theta$.}
 \label{fig:coordinates}
\end{figure}

But how well do rods with finite aspect ratios follow the Lagrangian stretching direction? In other words: how sensitive is the angular dynamics to small deviations $\delta\Lambda$ from $\Lambda=1$? To quantify this we compute
the steady-state distribution of the Euler angles $\phim$ and $\psim$ that quantify the angular separations between the orientation vector $\ve n$ and $\ns$ in the $\ns-\nc$ plane (yaw) and out of the $\ns-\nc$ plane (pitch), respectively,  as shown in Fig.~\ref{fig:coordinates}(a).

Near the channel center the flow is nearly isotropic, so it is not meaningful to ask how $\ve n(t)$ aligns with the basis vector of a laboratory-fixed frame. Near the wall, by contrast, the flow is anisotropic, and several authors have analysed how $\ve n(t)$ rotates w.r.t. the laboratory-fixed basis $\hat{\bf x}$ (flow direction), $\hat {\bf y}$ (span-wise direction), and $\hat{\bf z}$ (wall-normal direction). The results were summarized by \citet{Voth16}. Briefly, in the near wall turbulence the slender rods were found to preferentially align in the streamwise direction \citep{Mor08,Mar10} but thin disks tend to align in the wall-normal direction \citep{Zha15}. Euler angles $\theta$ and $\phi$ are defined respectively as the angle between the orientation vector $\ve n$ and the $\hat{ \bf x}-\hat{\bf z}$ plane, and the angle between $\ve n$ and $\hat{\bf x}$ in the $\hat{ \bf x}-\hat{\bf z}$ plane.

In conclusion there is a wealth of literature studying the alignment of nonspherical particles with local quantities as well as in a fixed frame. In this study we compute the precise effect of aspect ratio and the distance from the wall on alignment, and try to understand the underlying physical mechanisms.

\subsection{Direct numerical simulation of channel flow}
\label{sec:DNS}

We perform DNS of a turbulent channel flow at $Re_\tau = 180$. The friction Reynolds number is defined as $Re_\tau = h \, u_\tau/ \nu$, where $h$ is the channel half height, $u_\tau$ is the wall friction velocity, and $\nu$ is the kinematic viscosity of the fluid. We chose a domain of size $12h\times6h\times2h$ with 192 grid points in the streamwise ($\hat{\bf{x}}$), span-wise ($\hat{\bf{y}}$), and wall-normal ($\hat{\bf{z}}$) directions, respectively. Periodic boundary conditions are imposed in the homogeneous $\hat{\bf{x}}$ and $\hat{\bf{y}}$ directions and the no-slip and impermeability conditions are imposed at the walls. In the following, the subscript $+$ denotes normalization by the viscous scales, i.e. the viscous length scale $\nu/u_\tau$, and viscous time scale $\nu/u_\tau^2$.
The corresponding grid-resolution is uniform in the streamwise and spanwise directions, with grid spacings $\Delta x^+ = 11.3$ and $\Delta y^+ = 5.6$. The grid in the wall-normal direction is refined near the wall , and the spacing $\Delta z^+$ is 0.9 at the channel walls, but increases to 2.86 at channel center. A pseudo-spectral method is applied along the homogeneous directions and a second-order finite-difference discretisation is used in the wall-normal direction. Time integration is performed using a second-order explicit Adams-Bashforth scheme with time step $\Delta t^+= 0.036$ \citep{Zha16}.

Our simulations show nearly homogeneous and isotropic turbulence near the channel center \citep{And15}. Near the channel wall, however, the flow is anisotropic \citep{Man88, Pum17}. In particular, the no-slip boundary conditions induce a large mean shear near the wall and low/high speed streaks are formed in the near-wall region and have been observed in both experiment and numerical simulations \citep{Kim87}. The near wall turbulence structures and the presence of shear play important roles for inertial particle accumulation \citep{Mar02} and particle rotation \citep{Zha15}.
\section{DNS results} 
\label{sec:results}
\begin{figure}
	\centering
	\includegraphics[width=0.9\linewidth
		]{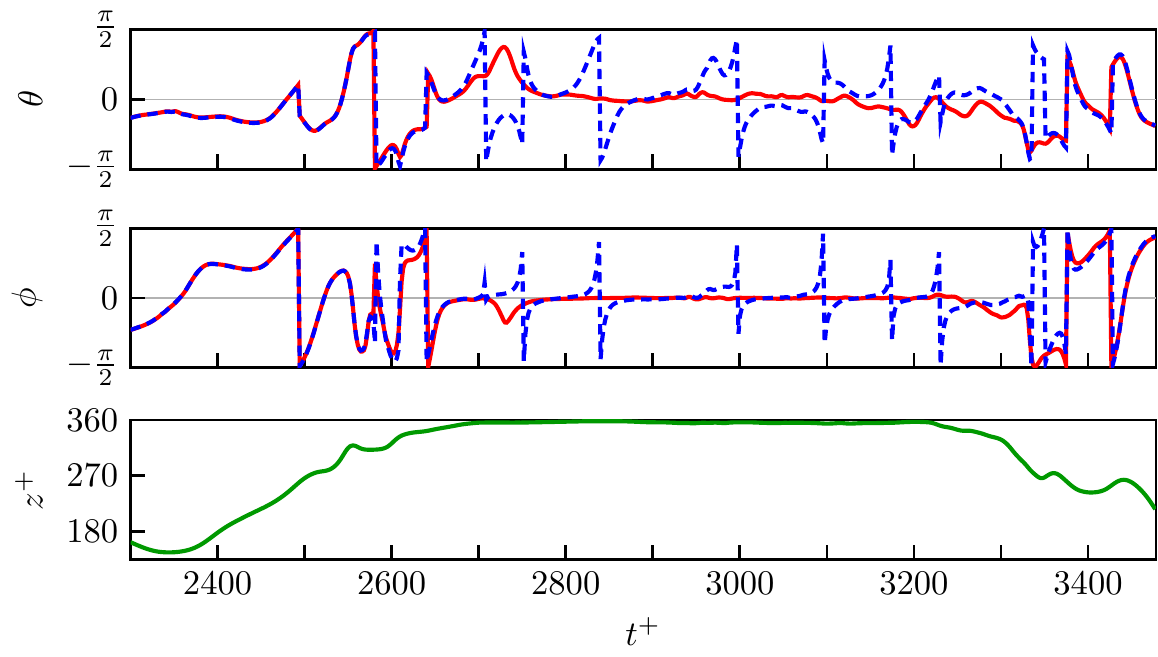}
\caption{Angular dynamics of a slender rod in the turbulent channel flow. The top two panels show the Euler angles $\theta$ and $ \phi$, see Fig.~\ref{fig:coordinates}(b), of the Lagrangian stretching direction $\ns(t)$ (solid red line), and of the orientation vector $\ve n(t)$  of a particle with $\Lambda = 0.9963$ (blue dashed line). The bottom panel shows the $z-$coordinate of the center-of-mass position of the particle in the channel. The channel boundaries are located at $z^+ = 0, 360$ with the centerline at $z^+=180$. During the time interval $2700<t^+<3200$ the particle travels in the viscous boundary layer, $354.5 <z^+<358.5$.}
	\label{fig:trajectory}
\end{figure}

\begin{figure}
	\centering	\vspace*{4mm}
	 \begin{overpic}[scale=1]{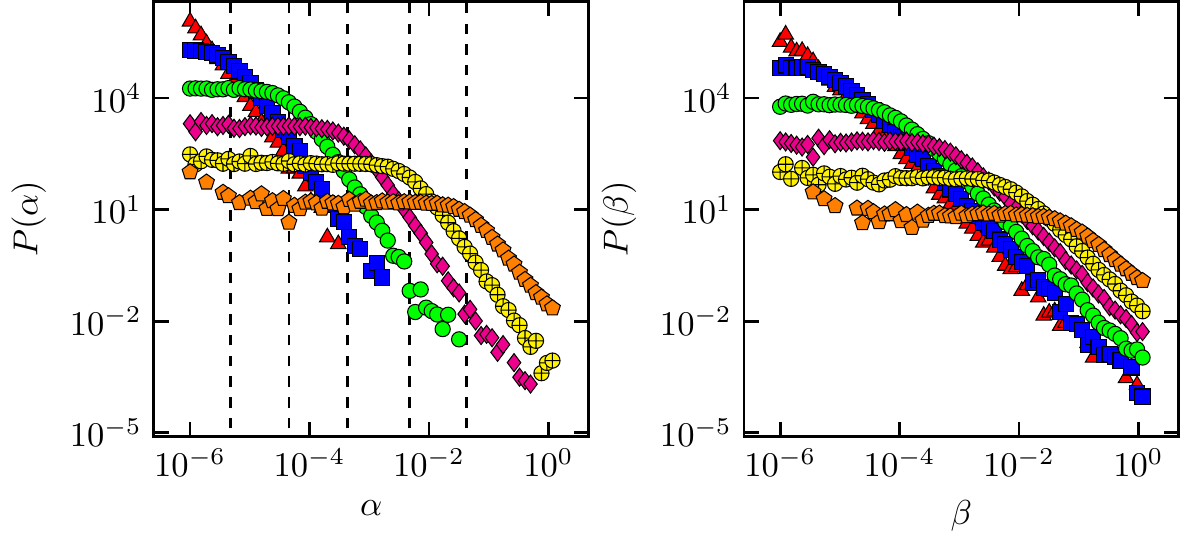}
	 	 \put(5,45){{(\textbf{a})}}
	 \put(55,45){{(\textbf{b})}}
	\end{overpic}
	\caption{Distribution of alignment between $\ve n(t)$ and $\ns(t)$ near the channel center at $z^+=180$. Shown are distributions of the Euler
	angles $\phim$ and $\psim$ , Fig.~\ref{fig:coordinates}(a). \textbf{Panel (a)}: distribution $P(\phim)$. Red triangle, blue square, green circle, magenta diamond, yellow $\oplus$, orange pentagon symbols correspond to $\delta\Lambda=10^{-6},10^{-5},10^{-4},10^{-3},10^{-2},10^{-1}$ respectively ($\delta \Lambda = 1-\Lambda$). Vertical dashed lines show the cutoff angles where the power law transitions to a plateau. \textbf{Panel (b)}: same but  for the distribution of $\psim$.}
	\label{fig:DNS_angle_distributions}
\end{figure}

\begin{figure}
	\centering	\vspace*{4mm}
	 \begin{overpic}[scale=1]{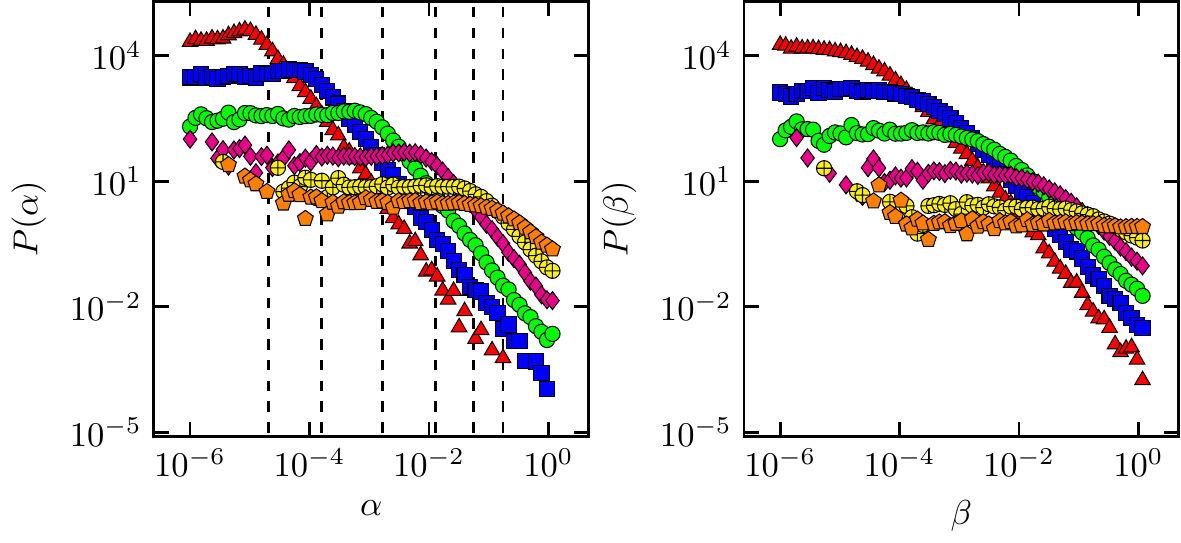}
	 \put(5,45){{(\textbf{a})}}
	 \put(55,45){{(\textbf{b})}}
	\end{overpic}
	\caption{Distribution of alignment between $\ve n(t)$ and $\ns(t)$ near the channel wall at $z^+ = 4$. Shown are distributions of the Euler
	angles $\phim$ and $\psim$ , Fig.~\ref{fig:coordinates}(a). \textbf{Panel (a)}: distribution $P(\phim)$. Red triangle, blue square, green circle, magenta diamond, yellow $\oplus$, orange pentagon symbols correspond to $\delta\Lambda=10^{-6},10^{-5},10^{-4},10^{-3},10^{-2},10^{-1}$ respectively ($\delta\Lambda = 1-\Lambda$). Vertical dashed lines show the cutoff angles where the power law transitions to a plateau. \textbf{Panel (b)}: same but  for the distribution of $\psim$.}
	\label{fig:DNS_angle_distributions_wall}
\end{figure}

\begin{figure}
	\centering
	 \begin{overpic}[scale=1]{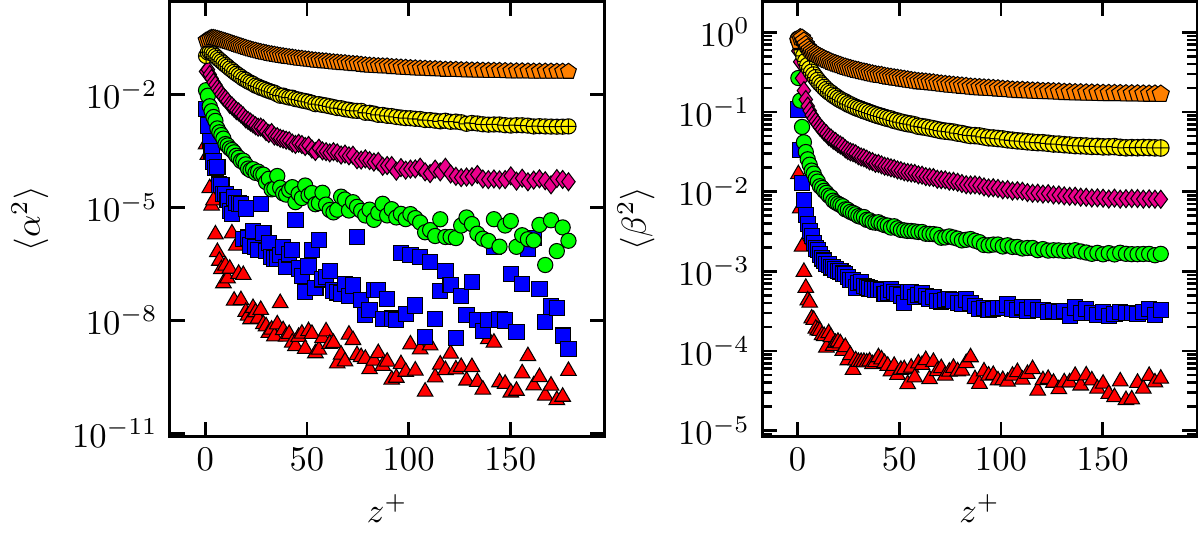}
	\end{overpic}
	\caption{The variances $\langle \alpha^2 \rangle, \langle \beta^2 \rangle$ of the angles $\alpha, \beta$ as a function of the wall normal distance $z^+$ for different values of $\delta \Lambda$. Red triangle, blue square, green circle, magenta diamond, yellow $\oplus$, orange pentagon symbols correspond to $\delta\Lambda=10^{-6},10^{-5},10^{-4},10^{-3},10^{-2},10^{-1}$ respectively ($\delta\Lambda = 1-\Lambda$).}
	\label{fig:cosbeta}
\end{figure}

Fig. ~\ref{fig:trajectory} shows DNS results for the angular dynamics of the orientation $\ve n(t)$ of a slender rod ($\Lambda = 0.9963$) in the channel flow, compared with the angular dynamics of the Lagrangian stretching direction $\ns(t)$.
Fig.~\ref{fig:trajectory} demonstrates that $\ve n(t)$ aligns quite well with the Lagrangian stretching direction when the particle is near the center of the channel. But when the particle is near the wall, then the angular dynamics of 
$\ve n(t)$ and $\ns(t)$ seem to be different: while $\ve n(t)$ tumbles in the near-wall shear flow, the Lagrangian stretching direction aligns with the streamwise direction $\hat{\bf x}$. Small deviations are due to  fluid-velocity fluctuations, but they do not seem to cause $\ns(t)$ to tumble. In order to investigate and quantify this difference, we compute the probability distributions of the angles $\phim$ and $\psim$ near the center (see Fig.~\ref{fig:DNS_angle_distributions}), and near the wall (see Fig.~\ref{fig:DNS_angle_distributions_wall}). The angles $\phim$ and $\psim$ characterise the difference between the particle orientation vector $\ve n(t)$ and the Lagrangian stretching direction $\ns(t)$. In particular, small $\phim, \psim$ correspond to good alignment between the particle orientation and the Lagrangian stretching direction. 
These Figures show that the relative angles near the channel wall are larger than near the channel center for the same shape parameter $\Lambda$.

In Fig.~\ref{fig:DNS_angle_distributions} we plot DNS results for the distribution of the Euler angles $\phim$ and $\psim$ near the channel center, at $z^+ = 180$.
 The figure shows distributions with power law tails that are cut off at small angles by a plateau in the distribution. The narrow plateaus at small values of $\phim$ and $\psim$ indicate that $\ve n(t)$ and $\ns(t)$ align well most of the time. But we also see that the distributions have power-law tails,
of the form $P(\alpha)\sim \alpha^{\xi}$.
For the largest values of $\Lambda$, close to unity, the exponent is close to $\xi=-2$, for both $\phim$ and $\psim$. For $\Lambda=0.9$ the exponents are slightly smaller, although the power laws are not as clear cut for this value
of $\Lambda$.

In Fig.~\ref{fig:DNS_angle_distributions_wall} we plot DNS results for the distribution of the angles $\phim$ and $\psim$ near the channel wall at $z^+ = 4$, characterising the difference between the particle-symmetry vector $\ve n(t)$ and the Lagrangian stretching direction
$\ns(t)$ (Fig.~\ref{fig:coordinates}a). The distributions look similar to the corresponding distributions near the channel center. One difference is that distributions near the channel wall have a broader plateau than the corresponding distributions near the channel center. This signifies larger relative angles between $\ve n(t)$ and $\ns(t)$. Both near the channel center and the channel wall, the distributions show a power law decay for large angles. This is an indication of large excursions in the relative angles between the orientation vector $\ve n(t)$ and $\ns(t)$.

In summary, we find that near the channel center rods with $\Lambda$ close to unity align well with the Lagrangian stretching direction $\ns(t)$. The distribution of the deviations
between $\ve n(t)$  and $\ns(t)$ has power-law tails that are cut off at small angles, giving rise to a plateau in the distribution. Near the channel wall, the alignment characteristics are similar: the distributions exhibit power-law tails at large relative angles and plateaus at small angles. However, the plateaus in the distributions near the channel wall are broader by an order of magnitude compared to the corresponding distributions near the center. This indicates that typically the fluctuations of relative angles are larger near the wall than near the center. The observed plateaus indicate random uncorrelated motion for small angles. We find that the mechanisms near the channel center and the channel wall are different. Near the channel center the distributions are a result of random fluctuations of the fluid-velocity gradients, whereas near the channel wall the dominant effects are the strong shear and the weak velocity-gradient fluctuations. In the next Section we discuss these observations and explain them using simple statistical models.
\section{Theory}
\label{sec:theory}

\subsection{Angular dynamics}

The angular dynamics of spheroidal particles in channel flows can be understood by looking at the equations of motion of the orientation vector $\ve n$ for $\Lambda =1$ and $\ve{n}_1$ for $\Lambda =1-\delta \Lambda$. Defining $\ve{\delta  n} = \ve{n}-\ve{n}_1$, the equations of motion for $\ve n, \ve{\delta  n}$ are,
\begin{subequations} \label{eq:angular}
\begin{align}
 \dot{\ve n} &= \ma A \ve n - (\ve n \cdot \ma S \ve n) \ve n , \\
 \dot{\ve {\delta n}} &= - \delta \Lambda [\ma S \ve n - (\ve n \cdot \ma S \ve n) \ve n] + [\ma A- (\ve n \cdot \ma S \ve n)] \ve {\delta n} - 2 (\ve {\delta n} \cdot \ma S \ve n) \ve n. \label{eq:deltan}
\end{align}
\end{subequations}
Here $\ma A$ is the fluid velocity gradient matrix and $\ma S$ is the symmetric part of $\ma A$. The first term in Eq.~(\ref{eq:angular}b) is independent of $\ve{\delta n}$, and the second term contains $\ve{\delta n}$ to the first power. These two terms can be thought of as being analogous to stochastic equations with additive and multiplicative noises.

\subsubsection{Angular dynamics near the channel center}
To qualitatively explain why the distributions in Fig.~\ref{fig:DNS_angle_distributions} have power laws we consider a two-dimensional model for the angular dynamics \citep{Zha19}.
 In two dimensions, the left Cauchy Green tensor has two eigenvectors, the expanding eigenvector $\ns$ and the contracting eigenvector $\nc$, but not the intermediate eigenvector $\nsc$. 
 This means that the two dimensional-model may explain the dynamics of the angle $\phim$ but not that of the angle $\psim$, Fig.~\ref{fig:coordinates}.  Following \citet{Gus16:rev} we model the homogeneous and isotropic fluid-velocity fluctuations as Gaussian random 
 functions that are white in time, but have smooth spatial correlations.
  In two dimensions, Eqs.~\eqref{eq:angular} can be written in terms of two angles: $\phi$, the angle that the Lagrangian stretching direction makes with the $x-$axis in the channel coordinates, and $\alpha \approx |\ve{\delta n}|, |\ve{\delta n}| \ll 1$, the angular separation between the particle and the Lagrangian stretching direction. The angular separation between the Lagrangian stretching direction and the symmetry vector $\ve n$ of a particle with shape parameter $\Lambda = 1-\delta\Lambda$ 
 is simply given by $\phim = \phi(\Lambda =1)-\phi(\Lambda =1-\delta\Lambda)$. Here $\phi_{\Lambda}$ follows Jeffery's equation Eq.~\eqref{eq:jeffery}, for $\ve n = [\cos \phi_\Lambda, \sin \phi_\Lambda]^{\sf T}$. 
  In the following, we drop the subscript in $\phi_{\Lambda=1}$. 
  
  We start off by assuming the fluid velocity-gradient matrix to be a Gaussian random variable with time correlation $\tau$. The dimensional parameters of the problem are the strength of velocity fluctuations, $u_0$, the correlation length $\eta$ and the correlation time $\tau$. Thus there are two relevant timescales, the correlation time $\tau$ and the advection time $\eta/u_0$. Out of these one can make one dimensionless parameter $\ku = u_0 \tau/ \eta$  \citep{Dun05}. We dedimensionalise the fluid-velocity gradient as $\ma A = \tfrac{\ku^2}{\tau} \ma A^\prime$ and the time as $t= \tfrac{\tau}{\ku^2} t^\prime$ and drop the primes in $\ma A, t$ in the following. Assuming traceless, and isotropy for the fluid gradient matrix $\ma A$, one finds that $\ma A$ has three independent components $O_{12}, S_{11},S_{12}$. Then for the variables $\phi$, and $\alpha$, Eqs.~\eqref{eq:angular} can be written as:
\begin{subequations}  \begin{align} \label{eq:2d_relative_angle}
 \dot{\phi} &=  - \sqrt{2}\, O_{12} -\sin 2 \phi \, S_{11} + \cos 2 \phi \, S_{12} ,\\
 \dot{\phim} &= -(2 \alpha \cos 2 \phi + \delta \Lambda \sin 2 \phi) \, S_{11} + (- 2 \alpha \sin 2 \phi + \delta \Lambda \cos 2 \phi) \,  S_{12}  \label{eq:2d_relative_angle_b}.
\end{align}
\end{subequations}
The white noise limit is taken as $\ku \to 0$, which corresponds to assuming that the fluid velocity correlation time is the shortest timescale in the problem, $\tau \ll \tfrac{\eta}{u_0}$. We find that the drift coefficients vanish. The diffusion coefficients are given by 
\begin{align} 
 \mathcal{D}_{\phim \phi} &=\mathcal{D}_{\phi \phim}= \tfrac{1}{2}  \, \delta\Lambda , \nonumber \\
 \mathcal{D}_{\phi \phi} &= \tfrac{3}{2}, \nonumber \\
 \mathcal{D}_{\phim  \phim} &= \tfrac{1}{2} \, (\delta \Lambda^2 + 4 \phim^2).\nonumber
\end{align}
{ The stationary Fokker-Planck equation for the joint distribution $P(\phim,\phi)$ is}
\begin{align}\label{eq:FP}
 \left[ \frac{3}{2}\frac{\partial^2}{\partial \phi^2} + \delta\Lambda \frac{\partial^2}{\partial \phi \ \partial \phim} +\frac{\partial^2}{\partial \phim^2} \frac{\delta\Lambda^2 + 4 \phim^2}{2}   \right]P(\phim,\phi) = 0\,.
\end{align}
We obtain the marginal distribution $P(\phim)$  by integrating out $\phi$ from Eq.~\eqref{eq:FP}, requiring symmetry $P(\phim) = P(-\phim)$,  and that $P(\phi)$ is normalized to unity. 
The result is:
\begin{align} \label{eq:prob_phim}
 P(\phim) = \frac{\delta \Lambda }{\Big[(\delta \Lambda) ^2+4 \phim ^2 \Big] \tan ^{-1}\left(\frac{\pi }{\delta \Lambda }\right)}.
\end{align}
Eq.~\eqref{eq:prob_phim} captures the qualitative features mentioned before, namely the power-law form of the distribution and its cut off at small angles, of the order of $\delta\Lambda$. Eq.~\eqref{eq:2d_relative_angle_b} shows that at small angles $\phim < \delta \Lambda$, $\dot{\phim}$ is dominated by the additive term, which is the term independent of $\phim$. This gives rise to the plateau in the distribution for $\phim$, indicating random uncorrelated motion of $\phi_\Lambda$ and the Lagrangian stretching direction.

Geometrically, the power law in $\alpha$ can be understood as a consequence of the fact that the Lagrangian stretching direction $e_{L1}(\ve x(t),t)$ acts as an attractor for the orientation field of particles with shape parameter $\Lambda$, $\phi_{\Lambda}(\ve x(t),t)$ along the same Lagrangian trajectory $\ve x(t)$. This clustering of orientations is analogous to spatial clustering of particles in turbulence in the advective limit \citep{Mei17,Gus16}, and correlated random walks in the inertia free limit \citep{Dub18}. Moreover, just as in the case of advected particles in one dimension, the power law in $\alpha$ is a result of purely diffusive dynamics with a diffusion coefficient proportional to $\alpha^2$, leading to the same power law exponent.

In the model the power-law exponent equals $-2$, and it is independent of $\delta\Lambda$. The exponent is very nearly $-2$ for the DNS results but since the theoretical model we have considered is two dimensional, we don't expect it to explain the exponent, but merely the mechanism. That the model agrees only qualitatively with DNS results is to be expected, the model neither describes the additional degree of freedom $\psim$, nor does it capture the persistent nature of the turbulent velocity-gradient fluctuations. But numerical simulations of a three-dimensional
model show qualitatively similar results.

\subsubsection{Angular dynamics near the channel wall}
\label{sec:theory_wall}
\begin{figure}
	\begin{minipage}{0.45\textwidth}
		\centering
		\begin{overpic}[scale=1]{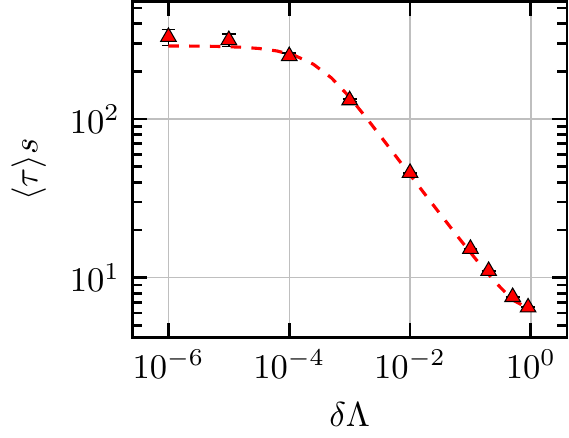}
			\put(25,72){{\textbf{a}}}
		\end{overpic}
	\end{minipage}\hfill
	\begin{minipage}{0.45\textwidth}
		\centering
		\begin{overpic}[scale=1]{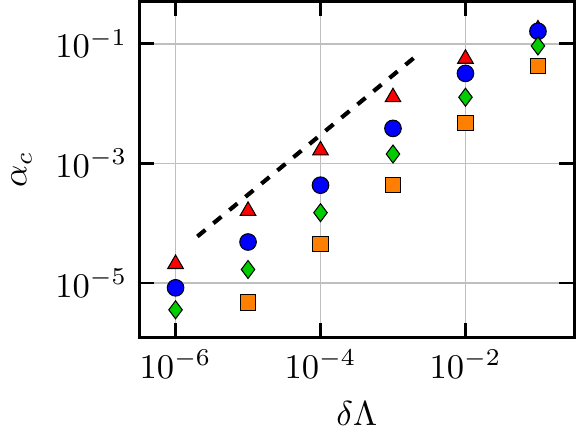}
			\put(25,70){{\textbf{b}}}
		\end{overpic}
	\end{minipage}
	
	\caption{Panel (\textbf{a}): Mean time between tumbles $\langle \tau \rangle s$ as a function of $\delta\Lambda = 1-\Lambda$ for a set of trajectories in the layer $0<z^+ <10$, $s$ is the mean shear along the observed trajectories. Symbols correspond to simulation results, the dashed line corresponds to theory.  Panel (\textbf{b}): $\alpha_{c}$ as a function of $\delta\Lambda$ for $z^+\sim 4$ (red triangles) and $z^+\sim 180$ (orange squares). $\alpha_c$ describes the transition between the power law and the plateau in Fig.~\ref{fig:DNS_angle_distributions}a in orange, and Fig.~\ref{fig:DNS_angle_distributions_wall}a in red. Blue circles and green diamonds correspond to $z^+ \sim 10, 20$ respectively. The black dashed line shows the reference slope $\delta  \Lambda ^1$, as predicted by theory.}
	\label{fig:tumbling_time}
\end{figure}

Near the channel wall the fluid-velocity gradient has a large shear component. Since the flow is anisotropic it is natural to express the angular dynamics in terms of
the channel-fixed basis $\hat{\bf x}$, $\hat{\bf y}$, and $\hat{\bf z}$ with the corresponding Euler angles $\theta$ and $\phi$, Fig.~\ref{fig:coordinates}(b).

We consider a simple model for a spheroidal particle experiencing simple shear and additive noise. \citet{Tur07} used this model to study the angular dynamics of single polymers in a chaotic
	flow with mean shear. While  \citet{Tur07} assumed $\Lambda=1$, here we consider general values of the shape parameter $\Lambda$.

Decompose the fluid gradient matrix $\ma A$ as a sum of the mean $\bar{\ma{A}}$ and the fluctuations $\ma{A}^\prime$ as $\ma A = \bar{\ma{A}} + \ma{A}^\prime$. The only non-zero component of $\bar{\ma{A}}$ is $\bar{A}_{xz} = s$ \citep{Cha15}.
Jeffery's equation Eq.~\eqref{eq:jeffery} for the angles $\phi,\theta$ in terms of $s$ and $\ma{A}^\prime$ is given by
\begin{align} 
 \dot{\phi} &= -\frac{s}{2} (1 -\Lambda \cos 2 \phi) + \eta_\phi, \label{eq:shear_noise}\\
 \dot{\theta} &= -\Lambda \, \frac{s}{4} \sin 2 \phi \sin 2 \theta + \eta_\theta. \label{eq:shear_noise_theta}
\end{align}
Here $\eta_\phi$ and $\eta_\theta$ are fluctuating terms which in general depend on $\ma{A}^\prime$, the angles $\theta, \phi$, and the shape parameter $\Lambda$. In the vicinity of the channel wall the shear strength $s$ is much larger than the magnitude of fluctuations of the velocity-gradient matrix, and thus the the particle spends long times close to $\phi = 0$. Since we have $\eta_\phi= A_{zx} + O(\phi)+ O(\theta)$ in the limit $\phi,\theta \to 0$, one can approximate $\eta_\phi \approx A_{zx}$. Let the autocorrelation for $\eta_\phi$ be given by $\langle\eta_\phi(0)\eta_\phi(t)\rangle = C_0 f(|t|/\tau)$, so that $C_0$ quantifies the magnitude of the fluctuations and $\tau$ is the correlation time. Then, Eq.~\eqref{eq:shear_noise} has three time scales, $1/s, 1/\sqrt{C_0}$ and $\tau$. Near the wall the shear strength is much larger than the strength of the fluctuations, $s \gg \sqrt{C_0}$. This gives that the dynamics of $\phi$ is comprised of two regions, the deterministic fast region with time scale $1/s$ and the stochastic slow region with time scale $1/\sqrt{C_0}$. Observation of typical trajectories show that near the channel wall the fluctuations of $\eta_\phi$ are much faster than the timescale of the slow, stochastic $\phi$ dynamics. Thus we take the white noise limit as $\tau \sqrt{C_0} \to 0$, while holding $2 D= \int_{-\infty}^\infty dt \, \eta_\phi(0) \eta_\phi(t) \propto C_0 \tau$ constant.

Fig.~\ref{fig:trajectory} shows a stark apparent difference between the dynamics of two particles, with slightly different shape parameters, $\Lambda =1$ and $\Lambda = 0.9963$. In particular one observes that near the wall, the Lagrangian stretching direction seems to align along $\hat{\bf x}$ whereas $\Lambda =0.9963$ seems to tumble along the same trajectory. This can be explained by calculating the mean time between tumbles, defined as the average time taken by the particle to travel from $\phi =+\pi/2$  to $\phi =-\pi/2$, Fig.~\ref{fig:tumbling_time}(a). A theoretical calculation using Eqs.~(\ref{eq:shear_noise}), see Appendix \ref{appendix:t_tumb}, shows good quantitative agreement with simulations. This is expected, considering that the parameter $\sqrt{C_0} \tau \sim 10^{-3}$ for the trajectory set used to obtain Fig.~\ref{fig:tumbling_time}(a), so that the white noise limit is a good approximation.

Next we analyse the dynamics for $\phi$. This is because the $\phi$ dynamics are independent of $\theta$, whereas the $\theta$ dynamics are slave to the process $\phi$.

Assuming $\eta_\phi$ to be a Gaussian random variable, white in time, with the intensity of fluctuations $2 D = \int_{-\infty}^\infty dt \, \eta_\phi(0) \eta_\phi(t)$, \citet{Tur07} obtained  the Fokker-Planck equation for the distribution of $\phi$,
\begin{align} \label{eq:fp_phi}
 \frac{\partial P(\phi,t)}{\partial t} = \frac{s}{2}\frac{\partial }{\partial \phi}\left[ (1 -\Lambda \cos 2 \phi)P(\phi,t)\right] + D \frac{\partial^2}{\partial \phi^2} P(\phi,t).
\end{align}

Eq.~\eqref{eq:fp_phi} differs from Eq.(9) in \citep{Tur07} because we consider general values of $\Lambda$ and not just $\Lambda =1$, and that the strength of fluctuations $D$ is defined slightly differently. Following \citet{Tur07}, the steady state distribution $P(\phi)$ is obtained as the time independent solution of Eq.~\eqref{eq:fp_phi},

\begin{align}
 P(\phi) = \mathcal{N} \int_0^\pi \mathrm{d}x \ \exp{-\frac{s}{2 D}(x -\Lambda \cos(2 \phi -x)\sin x)}.
\end{align}
Here one integration constant is determined by periodic boundary conditions, $P(\pi/2) = P(-\pi/2)$, and $\mathcal{N}$ is a normalization constant, which can be computed using $\int_{-\pi/2}^{\pi/2} \mathrm{d} \phi P(\phi) =1$.

In order to understand how the relative angle between a slender rod and the Lagrangian stretching direction behaves, next we analyse the relative angle, $\delta\phi = \phi(\Lambda =1)-\phi(\Lambda =1-\delta\Lambda)$. The joint equations of motion for $\phi, \delta\phi$ upto the second order in $\delta\phi$ are given by,

\begin{align} 
\dot{\phi} &= -\frac{s}{2} (1 -\cos 2 \phi) + \eta_\phi, \\
 \dot{\delta\phi} &= \frac{s}{2} \delta\Lambda \cos 2 \phi - (s \Lambda \sin 2\phi ) \delta\phi + (s \Lambda \cos 2\phi ) \delta\phi^2. \label{eq:relative_shear}
\end{align}
First assume $\phi$ takes its mean value, $\phi_0 \approx \tfrac{\sqrt{\pi}}{\Gamma(1/6)}\left(\tfrac{3}{2}\right)^{\tfrac{1}{3}} (D/s)^{\tfrac{1}{3}} \ll 1$ . Then we have $ \dot{\delta\phi} \approx \frac{s}{2} \delta\Lambda - (s \Lambda 2\phi_0 ) \delta\phi + (s \Lambda ) \delta\phi^2$ to the first order in $\phi_0$. This equation has fixed points $\phi_0 \pm \phi_0 \sqrt{1-\tfrac{\delta\Lambda}{2\Lambda \phi_0^2}}$. For $\delta\Lambda \ll 2 \Lambda \phi_0^2$ there is a stable fixed point at $\delta\phi_c \approx \tfrac{\delta\Lambda}{4 \Lambda \phi_0}$ and an unstable fixed point at $2 \phi_0 - \tfrac{\delta\Lambda}{4 \Lambda \phi_0}$. At the end of this section, we argue that near the channel wall the distributions of $\phim$ and $\psim$ get large contributions from the distributions of $\phi$ and $\theta$. Thus the stable fixed point explains the linear dependence of $\alpha_c$ on $\delta \Lambda$, Fig. \ref{fig:tumbling_time}(b). 

When $\delta\Lambda = 2 \Lambda \phi_0^2$, the two fixed points merge to one, and for $\delta\Lambda > 2 \Lambda \phi_0^2$ there are no fixed points. For  $\delta\Lambda > 2 \Lambda \phi_0^2$ the linear term in $\dot{\delta\phi}$ can be ignored.
 For the following we restrict ourselves to the regime $\delta\Lambda \ll 2 \Lambda \phi_0^2$ and ignore the $\delta\phi^2$ term in Eq. \eqref{eq:relative_shear}, 

\begin{equation}  \label{eq:relative_linear}
\begin{split}
\dot{\phi} &= -\frac{s}{2} (1 -\cos 2 \phi) + \eta_\phi, \\
 \dot{\delta\phi} &= \frac{s}{2} \delta\Lambda \cos 2 \phi - (s \Lambda \sin 2\phi ) \delta\phi.
\end{split}
\end{equation}

\citet{Che05} and \citet{Tur07} argued that the $\theta$ distribution must have power-law tails in the case of a single polymer in a shear flow, with different power-law exponents arising from deterministic and stochastic $\phi$ dynamics, but were unable to analytically estimate the value of the exponent for the stochastic region. Similarly, we find that the $\delta\phi$ distribution exhibits two regimes corresponding to deterministic and stochastic $\phi$ dynamics, both of which lead to power-law tails for the distribution of $\delta\phi$ with different exponents. In addition we calculate the power-law exponent both in the deterministic and the stochastic region. The deterministic regime $|\phi| \gg \phi_0$ leads to a power-law exponent  $-\tfrac{3}{2\Lambda}$ for $P(\delta\phi)$  and the stochastic regime $|\phi| \ll \phi_0$ with power law exponent  $-1-\tfrac{1}{\Lambda}$ for $P(\delta\phi)$.

First consider the deterministic regime, $|\phi| \gg \phi_0, \delta\phi > \tfrac{\delta\Lambda}{4 \Lambda \phi_0}$. In this regime the terms $\eta_\phi$ for $\dot{\phi}$ and $\tfrac{s}{2}\delta\Lambda \cos 2\phi$ for $\dot{\delta\phi}$ in Eqs.~\eqref{eq:relative_linear} can be ignored. These equations can then be integrated to obtain $\delta\phi = C \sin^{2\Lambda}\phi$ where $C$ is an integration constant. Using $P(\phi) \sim \phi^{-2}$, we obtain $P(\delta\phi) \sim \delta\phi^{-\tfrac{3}{2\Lambda}}$ by a change of variables.

Next consider the stochastic regime $|\phi| \ll \phi_0, \delta\phi > \tfrac{\delta\Lambda}{4 \Lambda \phi_0}$. Then the term $\tfrac{s}{2}\delta\Lambda \cos 2\phi$ in Eqs.~\eqref{eq:relative_linear} can be ignored. The steady state Fokker-Planck equation reads,

\begin{align}
 \frac{\partial }{\partial \phi}\left[ \sin^2 \phi  + \varepsilon^2 \frac{\partial}{\partial \phi}\right] P(\phi,\delta\phi) = - \Lambda \frac{\partial }{\partial \delta\phi}\left[ \sin 2 \phi \delta\phi P(\phi,\delta\phi)\right],
\end{align}

where we have defined $\varepsilon^2 = \tfrac{D}{2s}$ in analogy with \citet{Mei17}. Numerics suggests that the joint probability factorizes in the stochastic regime, so that $P(\phi, \delta\phi) = f(\phi) g(\delta\phi)$. We use separation of variables, and obtain $g(\delta\phi) \propto \delta\phi^{-1-\tfrac{\mu}{\Lambda}}$. The equation for $f(\phi)$ is a generalized eigenvalue problem,

\begin{align} \label{eq:phi_eigen}
 \frac{\partial }{\partial \phi}\left[ \sin^2 \phi  + \varepsilon^2 \frac{\partial}{\partial \phi}\right] f(\phi) = \mu \sin 2\phi \ f(\phi).
\end{align}
It is possible to obtain an eigenvector corresponding to the eigenvalue $\mu = 1$ for Eq.~\eqref{eq:phi_eigen}. The eigenvector has two undetermined integration constants which must be found by matching to the solution for large $\phi$, and normalization. Thus we conclude that $g(\delta\phi) = \delta\phi^{-1-\tfrac{1}{\Lambda}}$ is a solution for the tail of the distribution of $\delta\phi$ in the regime $|\phi| \ll \phi_0, \delta\phi > \tfrac{\delta\Lambda}{4 \Lambda \phi_0}$. When $\Lambda \approx 1$ this would lead to a power law for $\delta\phi$ with exponent $ \approx -2$.  In  Fig.~\ref{fig:DNS_angle_distributions_wall}(a) we have plotted the distribution for $\alpha$ which show a power law distribution for large $\alpha$ with exponent roughly $-2$. We argue next that since $\phim$ must be closely related to $\delta\phi$ near the channel wall, the observed exponent $-2$ for $\phim$ can be explained by our calculation for $\delta\phi$.

So far, we have analysed the dynamics for the angle $\phi$ in the channel frame, for particles with shape parameter $\Lambda$. Since the Lagrangian stretching direction spends long times aligned with $\hat{\bf{x}}$ and the Lagrangian contracting direction spends long times aligned with $\hat{\bf{z}}$, $\alpha, \beta$ get large contributions from $\delta\phi, \delta\theta$ respectively. This means that near the channel wall one can use $\delta \phi$ and $\delta \theta$ as a proxy for understanding $\alpha$ and $\beta$ respectively. The precise calculation of the distributions of $\alpha, \beta$ is an open question left for future work. 

In the presence of strong mean shear and weak velocity-gradient fluctuations we have shown that the width of the plateau in the distribution of relative angles scales linearly with $\delta\Lambda$ and that the power-law exponent for the tail of the distribution is $-\tfrac{3}{2\Lambda}$ and $-1-\tfrac{1}{\Lambda}$ in the deterministic and stochastic regimes respectively. Consider the relative angle in the case of constant shear without fluctuations. In the long time, the Lagrangian stretching direction reaches a steady state $\phi=0$, and thus the distribution of $\delta\phi$ has a plateau whose width scales as $\delta\Lambda^{\tfrac{1}{2}}$ and an exponent $-2$ for the power-law tail of the distribution. Thus, the presence of weak fluctuations affects the relative angular dynamics sensitively and the observations cannot be explained in terms of just the strong mean shear.

\section{Discussion} \label{sec:discussion}
We have seen that the alignment of slender rods with the Lagrangian stretching direction in a channel flow depends on the distance of the particle from the channel wall, as well as the particle shape parameter $\Lambda$. We found that the distributions of relative angles have power-law tails. A power-law tail implies that the relative angle exhibits large excursions. The power law tails are cut off at small angles by a plateau. A plateau in the distribution implies that the relative angles are essentially uncorrelated at small angles. Near the channel center the power law is a result of purely diffusive dynamics for the relative angle. Near the channel wall, by contrast, the power-law exponent is a result of the stochastic Lagrangian stretching direction dynamics and thus is a consequence of weak velocity-gradient fluctuations in addition to the strong mean shear. The plateau is broader near the channel wall than near the channel center for particles with the same shape parameter. The width of the plateau near the channel wall depends on the ratio of the mean shear to the fluctuation strength of the velocity-gradient matrix element $A_{zx}$. Since the variance of the relative angle is described by the width of the plateau in the distribution, the large relative angles near the wall are a result of the mean shear strength being much larger than the velocity-gradient fluctuations. In general this implies that the particles show better alignment near the channel center compared to near the channel wall. 

A related important problem is understanding the relative angle between two particles as they approach each other. Since the velocity field is smooth at small scales, one might expect that as non-spherical particles approach each other,  and tend to align in the same direction. However, \citet{Zha19} found that the relative angles between non-spherical particles close to each other show large excursions away from zero. This is quantified by the angular structure functions. The angular structure functions, $\langle |\psi(r)|^p \rangle$, of spheroidal particles are a measure of the relative orientations of two particles with the same shape parameter $\Lambda$ at distance $r$. For homogeneous isotropic turbulence, the structure functions were studied by \citet{Zha19}. The problem we have considered in this study, the relative angular orientation of a rod-like particle with respect to the Lagrangian stretching direction is closely related to the structure functions.

Firstly, the problem of understanding the angular structure functions can be broken down into two parts: (a) understanding how a particle with shape parameter $\Lambda$ aligns with respect to the Lagrangian stretching direction, which is a unique local reference vector, and (b) how the Lagrangian stretching directions at a spatial separation $r$ align with respect to each other. In this article we have tackled problem (a). Secondly, in two dimensions, the equations of motion for the angular structure functions, Eqs. (A1a)-(A1c) in \citet{Zha19}, exhibit striking similarities to the equations of motion for the problem discussed in this article (Eq.~\ref{eq:2d_relative_angle}), the angular separation of a particle with shape parameter $\Lambda$ and the Lagrangian stretching direction. 

\begin{figure}
	\centering
	 \begin{overpic}[scale=1]{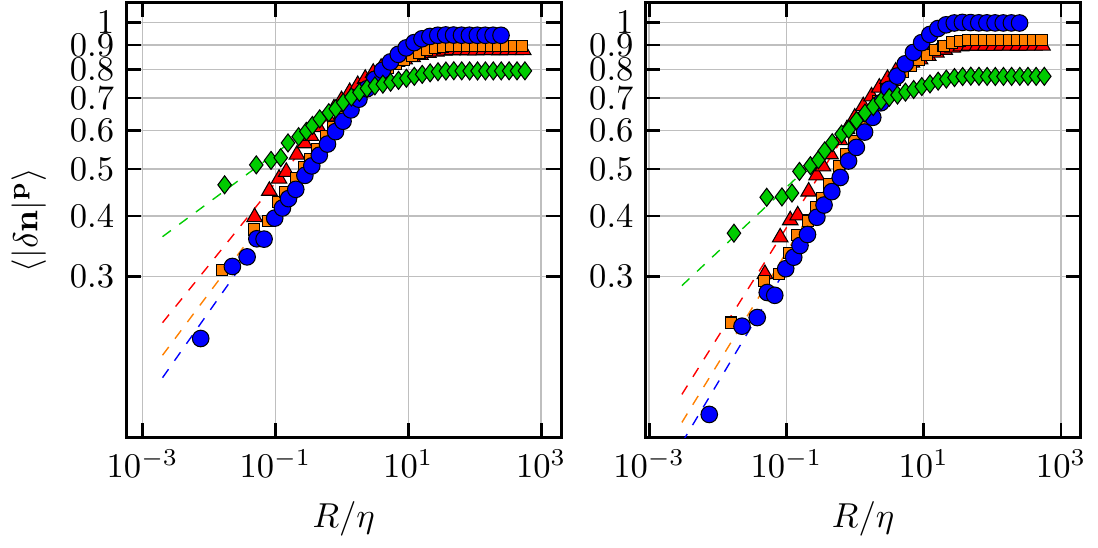}
		\put(21,40){$p=1$}
		\put(68,40){$p=2$}
	\end{overpic}
	\caption{Angular structure functions $S_p(r)$ plotted as a function of the particle separation $r$ for $\Lambda = 0.8$. Green diamonds, red triangles, orange squares, blue circles correspond to $z^+ \sim 0-5, 5-15, 15-25, 175-185,$ respectively. The left, and right panels shows the structure functions for $p=1$, and $p=2$, respectively. The dashed lines show power law fits in the limit $R/\eta \to 0$. Observe that the scaling behavior in the viscous layer $z^+<5$ is different from the scaling in the buffer layer, and near the center of the channel.
	}
	\label{fig:structurefn}
\end{figure}

In particular, the two equations for the angular separations have the same multiplicative term $(\ve n \cdot \ma S\ve n)$. The two equations have different additive terms where the cutoff in Eq.~\eqref{eq:2d_relative_angle} is set by the small value $\delta\Lambda$ and the cutoff in the case of \citet{Zha19} is set by the small separation $\ve R$. In general, however, the required distribution of $\delta\psi$ is conditioned on small $\ve R$, which might in turn modify the distribution of the multiplicative term $\ve n \cdot \ma S \ve n$.   However, the calculation of this precise distribution remains an open problem.

In a channel flow, one observes different scalings for the angular structure functions near the wall, in the viscous layer ($z^+ < 5$), compared to the rest of the channel, Fig.~\ref{fig:structurefn}. The smaller value of the scaling exponent near the wall implies larger relative angles between particles compared to the rest of the channel. Physically, we expect the mechanism causing larger relative angles between nearby particles near the channel wall to be related to the larger relative angles between the passive directors and the local Lagrangian stretching direction, however, the precise calculation is left for future work.
\section{Conclusions}
\label{sec:conclusions}

We investigated the alignment of slender rod-like particles with the local Lagrangian stretching direction in a turbulent channel flow. Our goal was to understand how this alignment depends on the distance of the particle from the channel wall and upon the particle shape parameter $\Lambda$. To this end we calculated the distribution of the relative angles between a particle with shape parameter $\Lambda = 1- \delta\Lambda$ and the Lagrangian stretching direction for different particle shape parameters and distances from the channel wall. We found that the distribution of relative angles exhibits power-law tails at large angles cut off by plateaus at small angles both near the channel center and the channel wall. At small angular separations the relative angles essentially perform uncorrelated random motion, leading to plateaus in the relative angle distributions. However, the relative angle exhibits rare large excursions. These large excursions lead to power-law tails in the distributions of relative angles. The variance of the relative angle is closely related to the width of the plateau which depends on the distance of the particle from the channel wall as well as the shape parameter. We found that the width of the plateau is proportional to $\delta\Lambda$ both near the channel center and near the channel wall. However, the plateau near the channel wall is much broader than near the channel center. Thus the alignment of a slender rod with the Lagrangian stretching direction is stronger near the channel center than near the channel wall.

We explained these observations using simple statistical models based on Jeffery's equation. Near the channel center where turbulence is approximately homogeneous and isotropic, we used a two dimensional toy model to qualitatively understand the distribution of relative angles. Near the channel wall, where the turbulent velocity-gradient fluctuations are small and the mean shear rate is large, we used the diffusion approximation for the three dimensional dynamics to find excellent agreement with numerical simulations. The diffusion approximation works quantitatively near the channel wall because for very slender particles and for the Lagrangian stretching direction, the timescale of angular dynamics is much slower than the timescale of fluctuations of the velocity-gradient term $A_{zx}$, which acts as additive noise. Mathematically, the fact that the distributions of relative angles are plateaus for small angles followed by power-law tails at large angles can be seen as a consequence of the general structure of the equations for relative orientations, Eq.~\eqref{eq:deltan}, which is analogous to that of a multicomponent stochastic process with additive and multiplicative noise \citep{Deu94}. The additive term gives rise to the plateau whereas the multiplicative term gives rise to the power-law tail. The plateau followed by a power-law tail in the distribution indicates that the particles spend most of the time at small relative angles, performing uncorrelated random motion, but rarely the relative angles show large excursions away from alignment. The power-law exponent for the tails of the relative angle distributions quantifies the frequency of the excursions. Near the channel center, the power-law tails are a result of diffusive relative angular dynamics, with a diffusion coefficient that increases with increasing relative angle. This is analogous to the relative separation of advected particles in turbulence \citep{Mei17, Gus16}. On the other hand near the channel wall the power-law tails are a result of the weak velocity-gradient fluctuations and strong mean shear. This is because the dynamics of the relative angle between a very slender particle's orientation and the Lagrangian stretching direction depends sensitively on the dynamics of the Lagrangian stretching direction, which in turn depends on the strong mean shear and the weak fluctuations of the velocity-gradient matrix element $A_{zx}$. The equation of motion of the Lagrangian stretching direction in the long time limit is the same as that of a single infinitely slender polymer, whose dynamics in strong mean shear with weak isotropic fluctuations was analysed by \citet{Tur07}. The variance of the relative angle is given by the width of the plateau of the distribution. In our two-dimensional toy model the width of the plateau is of the order of $\delta\Lambda$ near the center, but the model near the channel wall predicts a width of the order of $\delta\Lambda (s/D)^{1/3}$. Here $s$ is the strength of the mean shear and $D$ corresponds to the strength of fluctuations of the fluid velocity-gradient matrix element $A_{zx}$. Since $s/D \gg 1$, the plateau is broader near the channel wall, which explains the large relative angles between particle orientation and the Lagrangian stretching direction near the channel wall. Thus the large relative angles observed near the channel wall are a consequence of both the weak fluctuations of the fluid velocity-gradient element $A_{zx}$ and the strong mean shear. The importance of weak velocity-gradient fluctuations can be seen by the fact that in the absence of fluctuations, when the velocity-gradient matrix is constant with only a shear component, the width of the plateau scales as $\delta\Lambda^{\tfrac{1}{2}}$ instead of $\delta\Lambda$ as observed.

The alignment of elongated particles along streamlines is important for the paper making industry \citep{Car10} as well as bacteria density profiles in inhomogeneous flows \citep{Deh19}. Our results near the channel wall also explain how this orientation depends on the shape parameter of the particle. Another important question for the paper making industry is the relative alignment of two nearby fibres. We have argued that this problem of understanding the relative angle distribution between two nearby elongated particles is related to the problem we have considered in the present study, and leave the calculation of the relative angle between nearby particles for arbitrary shape parameter for future work. Further, it would be interesting to analyse the relative angular dynamics between thin discs and the Lagrangian contracting direction. We expect the distributions of relative angles to also exhibit plateaus for small angles and power-law  tails for large angles.

\begin{acknowledgments}
	This work was supported in part by  Vetenskapsr\aa{}det
	(grant number 2017-03865), by the grant \lq Bottlenecks for particle growth in turbulent aerosols\rq{} from the Knut and Alice Wallenberg Foundation, Dnr. KAW 2014.0048,
	and by a collaboration grant from the joint China-Sweden mobility programme (NSFC-STINT) [grant numbers 11911530141 (China), CH2018-7737 (Sweden)]. ZC and LZ are grateful for the support of the Natural Science Foundation of China through Grant 11702158, 91752205 and computational resources (grant number NN2649K) from The Research Council of Norway. AD would like to thank Kristian Gustavsson and Jan Meibohm for helpful discussions.
\end{acknowledgments}

\appendix
\section{Time between subsequent tumbles} \label{appendix:t_tumb}
In order to explain the observed difference in tumbling characteristics for particles with $\Lambda =1$ and $\Lambda =0.9963$, Fig.~\ref{fig:trajectory}, we calculate the mean time between subsequent tumbles as a function of the shape parameter $\Lambda$. \citet{Deh19} performed the same calculation for shape parameter $\Lambda =1$. \citet{Tur07} calculated the mean tumbling frequency and the distribution of times between subsequent tumbles, both for $\Lambda =1$. We define the time between subsequent tumbles as the time it takes for particles to travel from $\phi = +\pi/2$ to $\phi = -\pi/2$. To this end we start with Eq.~\eqref{eq:shear_noise} for $\phi$,
\begin{align} \label{eq:t_tumb_phi}
 \dot{\phi} &= -\frac{s}{2} (1 -\Lambda \cos 2 \phi) + \eta_\phi.
\end{align}
Here $\eta_\phi$ is a Gaussian random variable, with $\langle \eta_\phi(t)\rangle =0, \langle \eta_\phi(t)\eta_\phi(t^\prime)\rangle = 2 D\delta(t-t^\prime)$. The corresponding Fokker-Planck equation can be written,

\begin{align}
\frac{\partial P(\phi,t)}{\partial t} = \mathcal{L}_{FP} P(\phi,t), \\
 \mathcal{L}_{FP}(\phi) = D \frac{\partial }{\partial \phi} e^{-\tfrac{f(\phi)}{D}} \frac{\partial }{\partial \phi} e^{\tfrac{f(\phi)}{D}},
\end{align}
where $f(\phi) = \frac{s}{2} (\phi-\frac{\Lambda}{2}\sin 2\phi)$. Then the mean exit time $T_1(\phi^\prime)$ to exit the domain $\Omega = [\pi/2,-\pi/2)$ starting at $\phi^\prime = \pi/2$ can be calculated as follows. Let $P(\phi,t| \phi^\prime,0)$ be the transition probability from $ \phi^\prime$ at time $0$ to $\phi$ at time $t$. The initial condition for the transition probability is $P(\phi,0|\phi^\prime,0) = \delta(\phi-\phi^\prime)$. The probability that a trajectory starting at $\phi^\prime \in \Omega$ at time $0$ is still in the domain $\Omega$ at time $t$ is $\int_\Omega \mathrm{d} \phi P(\phi,t|\phi^\prime,0)$. This gives $1 - \int_\Omega \mathrm{d}\phi P(\phi,t|\phi^\prime,0)$ as the cumulative probability that the first exit time is greater than $t$. This means that the probability density for the first exit time $\rho(t)$ is given by 

\begin{align}
 \rho(t) = -\int_\Omega \mathrm{d}\phi \frac{\partial}{\partial t} P(\phi,t|\phi^\prime,0).
\end{align}
Thus one obtains the mean first exit time as,

\begin{align}
 T_1(\phi^\prime) = \int_0^\infty \mathrm{d}t \ t \rho(t) =  -\int_\Omega \mathrm{d}\phi \int_0^\infty \mathrm{d}t \ t \frac{\partial}{\partial t} P(\phi,t|\phi^\prime,0).
\end{align}
Defining $p_1(\phi,\phi^\prime) = -\int_0^\infty \mathrm{d}t \ t \frac{\partial}{\partial t} P(\phi,t|\phi^\prime,0)$, we obtain by integration by parts, $p_1(\phi,\phi^\prime) = \int_0^\infty \mathrm{d}t \ P(\phi,t|\phi^\prime,0)$. Then we have,

\begin{align}
 \mathcal{L}_{FP} p_1(\phi,\phi^\prime) = \int_0^\infty \mathrm{d}t \frac{\partial}{\partial t}\ P(\phi,t|\phi^\prime,0) = - \delta(\phi-\phi^\prime).
\end{align}
Thus $p_1(\phi,\phi^\prime)$ satisfies the differential equation,

\begin{align} \label{eq:tumble_time}
 \mathcal{L}_{FP} p_1(\phi,\phi^\prime) = -\delta(\phi-\phi^\prime),
\end{align}
where $p_1(\phi,\phi^\prime)$ must satisfy the same boundary conditions as $P(\phi,t|\phi^\prime,0)$. For the calculation of the mean time it takes a trajectory starting at $\phi^\prime = \pi/2 - 0$ to travel to $-\pi/2$ we use reflecting boundary condition at $\pi/2$ and absorbing boundary condition at $-\pi/2$. Then, Eq.~\eqref{eq:tumble_time} can be integrated to give,

\begin{align}
 p_1(\phi,\pi/2-0) = - \frac{1}{D} e^{-\tfrac{f(\phi)}{D}} \int_{-\pi/2}^\phi dy e^{\tfrac{f(\phi)}{D}} \int_{\pi/2}^y \mathrm{d}z \delta(z-\pi/2 + 0).
\end{align}
This gives for the mean time between tumbles,

 \begin{align}
 T_1(\pi/2-0) = \int_{\pi/2}^{-\pi/2} \mathrm{d}\phi \ p_1(\phi,\pi/2-0) = \frac{1}{D} \int_{-\pi/2}^{\pi/2}  \mathrm{d}\phi  \ e^{-\tfrac{f(\phi)}{D}} \int_{-\pi/2}^\phi \mathrm{d}y \ e^{\tfrac{f(y)}{D}}.
\end{align}
The corresponding equation in \citep{Deh19} is SI Appendix, Eq.(19), with $\Lambda =1$ and different boundary conditions. We further simplify this expression and obtain the asymptotics in the limit of small and large $\delta\Lambda$. Physically this corresponds to the transition in the shape parameter where the additive noise $\eta_\phi$ is no longer important. We obtain,
\begin{align}
 T_1 =  \frac{1}{D} \int_{-\pi/2}^{\pi/2}  \mathrm{d}\phi  \int_{-\pi/2-\phi}^0 \mathrm{d}y \ e^{\tfrac{s}{2D}(y- \Lambda \cos (y+2\phi) \sin y)}
\end{align}
when $\tfrac{s}{D} \gg 1$, the inner integral gets a large contribution near $y=0$, and decays quickly away from $y=0$, therefore we replace the lower limit in the inner integral by $-\infty$,then one can perform the integral over $\phi$ to obtain
\begin{align}
 T_1 &=  \frac{\pi}{D} \int_{-\infty}^0 \mathrm{d}y \ e^{\tfrac{s}{2D}y + \log I_0( \tfrac{s}{2 D} \Lambda \sin y)},\\
 &=  \frac{\pi}{D} \int_{0}^\infty \mathrm{d}y \ e^{-\tfrac{s}{2D}y + \log I_0( \tfrac{s}{2 D}\Lambda \sin y)}, \\
\end{align}
where we have changed integration variables $y \to -y$, and $I_0(z)$ is the modified Bessel function. Next we use the asymptotic expansion of the modified Bessel function for large argument, $\log I_0(z) \sim z -\tfrac{1}{2}\log (2 \pi z)$ to obtain
\begin{align}
 T_1 = \frac{\pi}{D} \int_{0}^\infty \mathrm{d}y \ e^{-\tfrac{s}{2D}y + \tfrac{s}{2 D}\Lambda \sin y -\tfrac{1}{2} \log\left( 2 \pi  \tfrac{s}{2 D}\Lambda \sin y\right)}.
\end{align}
Using the Taylor expansion of the sine function we get,

\begin{align}
 T_1 = \frac{\pi}{D} \int_{0}^\infty \mathrm{d}y \ e^{-\tfrac{s}{2D}y + \tfrac{s}{2 D}\Lambda (y-\tfrac{y^3}{6}) -\tfrac{1}{2} \log\left( 2 \pi  \tfrac{s}{2 D}\Lambda y\right)}. \\
\end{align}
Thus we have,

\begin{align}
 T_1 = \frac{\sqrt{\pi} 12^{\tfrac{1}{6}}}{D} \left(\frac{D}{s \Lambda}\right)^{\tfrac{2}{3}} \int_{0}^\infty \mathrm{d}y  \ \frac{1}{\sqrt{y}} e^{- \left( \tfrac{3}{2}\right)^{\tfrac{1}{3}}\left(\tfrac{s}{D}\right)^{\tfrac{2}{3}} \tfrac{\delta\Lambda}{\Lambda^{1/3}} y - y^3 }. \\
\end{align}
The behavior of the integrand changes depending on the magnitude of the coefficient of $y$ in the integrand. When $\delta \Lambda \ll \left(\tfrac{D}{s}\right)^{\tfrac{2}{3}}$ the linear term can be neglected and approximating $\Lambda \approx 1$ the result is,
\begin{align} \label{eq:tumb_rod}
 T_1 = \frac{\sqrt{\pi}2^{\tfrac{4}{3}} 3^{\tfrac{1}{6}} \Gamma(\tfrac{7}{6})}{D} \left(\frac{D}{s}\right)^{\tfrac{2}{3}}.
\end{align}
On the other hand when $\delta \Lambda \gg \left(\tfrac{D}{s}\right)^{\tfrac{2}{3}}$, the linear term dominates over the cubic term, and the result is,
\begin{align} \label{eq:tumb_rod}
 T_1 = \frac{\sqrt{2} \pi}{s} \delta\Lambda^{-\tfrac{1}{2}}, 
\end{align}
as predicted by Jeffery's theory.


\clearpage


\begin{thebibliography}{48}
\expandafter\ifx\csname natexlab\endcsname\relax\def\natexlab#1{#1}\fi
\def\au#1{#1} \def\ed#1{#1} \def\yr#1{#1}\def\at#1{#1}\def\jt#1{\textit{#1}}
  \def\bt#1{#1}\def\bvol#1{\textbf{#1}} \def\vol#1{#1} \def\pg#1{#1}
  \def\publ#1{#1}\def\arxiv#1{#1}\def\org#1{#1}\def\st#1{\textit{#1}}

\bibitem[Andersson {\em et~al.\/}(2015)Andersson, Zhao \& Variano]{And15}
{\sc \au{Andersson, Helge~I.}, \au{Zhao, Lihao} \& \au{Variano, Evan~A.}}
  \yr{2015}  \at{On the {Anisotropic} {Vorticity} in {Turbulent} {Channel}
  {Flows}}.  \jt{J. Fluids Eng}  \bvol{137},  \pg{084503--084503--3}.

\bibitem[Balkovsky \& Fouxon(1999)]{Bal99}
{\sc \au{Balkovsky, E.} \& \au{Fouxon, A.}} \yr{1999}  \at{Universal long-time
  properties of lagrangian statistics in the batchelor regime and their
  application to the passive scalar problem}.  \jt{Phys. Rev. E}  \bvol{60},
  \pg{4164--4174}.

\bibitem[Bezuglyy {\em et~al.\/}(2010)Bezuglyy, Mehlig \& Wilkinson]{Wil10a}
{\sc \au{Bezuglyy, V.}, \au{Mehlig, B.} \& \au{Wilkinson, M.}} \yr{2010}
  \at{Poincar\'e indices of rheoscopic visualisations}.  \jt{Europhys. Lett.}
  \bvol{89},  \pg{34003}.

\bibitem[Borgnino {\em et~al.\/}(2019)Borgnino, Gustavsson, De~Lillo, Boffetta,
  Cencini \& Mehlig]{Bor19}
{\sc \au{Borgnino, M.}, \au{Gustavsson, K.}, \au{De~Lillo, F.}, \au{Boffetta,
  G.}, \au{Cencini, M.} \& \au{Mehlig, B.}} \yr{2019}  \at{Alignment of
  nonspherical active particles in chaotic flows}.  \jt{Phys. Rev. Lett.}
  \bvol{123},  \pg{138003}.

\bibitem[Bretherton(1962)]{Bretherton:1962}
{\sc \au{Bretherton, F.P.}} \yr{1962}  \at{The motion of rigid particles in a
  shear flow at low {R}eynolds number}.  \jt{J. Fluid Mech.}  \bvol{14},
  \pg{284--304}.

\bibitem[Byron {\em et~al.\/}(2015)Byron, Einarsson, Gustavsson, Voth, Mehlig
  \& Variano]{Byron2015}
{\sc \au{Byron, M.}, \au{Einarsson, J.}, \au{Gustavsson, K.}, \au{Voth, G.},
  \au{Mehlig, B.} \& \au{Variano, E.}} \yr{2015}  \at{Shape-dependence of
  particle rotation in isotropic turbulence}.  \jt{Phys. Fluids}  \bvol{27},
  \pg{035101}.

\bibitem[Carlsson {\em et~al.\/}(2010)Carlsson, S{\"o}derberg \&
  Lundell]{Car10}
{\sc \au{Carlsson, Allan}, \au{S{\"o}derberg, L.~D.} \& \au{Lundell, Fredrik}}
  \yr{2010}  \at{Fibre orientation measurements near a headbox wall}.
  \jt{Nordic Pulp \& Paper Research Journal}  \bvol{25},  \pg{204--212}, qC
  20140901.

\bibitem[Challabotla {\em et~al.\/}(2015)Challabotla, Zhao \& Andersson]{Cha15}
{\sc \au{Challabotla, Niranjan~Reddy}, \au{Zhao, Lihao} \& \au{Andersson,
  Helge~I.}} \yr{2015}  \at{Shape effects on dynamics of inertia-free spheroids
  in wall turbulence}.  \jt{Phys. Fluids}  \bvol{27},  \pg{061703}.

\bibitem[Chertkov {\em et~al.\/}(2005)Chertkov, Kolokolov, Lebedev \&
  Turitsyn]{Che05}
{\sc \au{Chertkov, M.}, \au{Kolokolov, I.}, \au{Lebedev, V.} \& \au{Turitsyn,
  K.}} \yr{2005}  \at{Polymer statistics in a random flow with mean shear}.
  \jt{J. Fluid Mech.}  \bvol{531},  \pg{251–260}.

\bibitem[Chevillard \& Meneveau(2013)]{Che13}
{\sc \au{Chevillard, L.} \& \au{Meneveau, C.}} \yr{2013}  \at{Orientation
  dynamics of small, triaxial-ellipsoidal particles in isotropic turbulence}.
  \jt{J. Fluid Mech.}  \bvol{737},  \pg{571}.

\bibitem[Dehkharghani {\em et~al.\/}(2019)Dehkharghani, Waisbord, Dunkel \&
  Guasto]{Deh19}
{\sc \au{Dehkharghani, Amin}, \au{Waisbord, Nicolas}, \au{Dunkel, J{\"o}rn} \&
  \au{Guasto, Jeffrey~S.}} \yr{2019}  \at{Bacterial scattering in microfluidic
  crystal flows reveals giant active taylor{\textendash}aris dispersion}.
  \jt{Proceedings of the National Academy of Sciences}  \bvol{116},
  \pg{11119--11124}.

\bibitem[Deutsch(1994)]{Deu94}
{\sc \au{Deutsch, J.M.}} \yr{1994}  \at{Probability distributions for
  multicomponent systems with multiplicative noise}.  \jt{Physica A Statistical
  Mechanics and its Applications}  \bvol{208}~(3),  \pg{445--461}.

\bibitem[Dubey {\em et~al.\/}(2018)Dubey, Meibohm, Gustavsson \& Mehlig]{Dub18}
{\sc \au{Dubey, A.}, \au{Meibohm, J.}, \au{Gustavsson, K.} \& \au{Mehlig, B.}}
  \yr{2018}  \at{Fractal dimensions and trajectory crossings in correlated
  random walks}.  \jt{Phys. Rev. E}  \bvol{98},  \pg{062117}.

\bibitem[Duncan {\em et~al.\/}(2005)Duncan, Mehlig, \"Ostlund \&
  Wilkinson]{Dun05}
{\sc \au{Duncan, Kevin}, \au{Mehlig, Bernhard}, \au{\"Ostlund, Stellan} \&
  \au{Wilkinson, Michael}} \yr{2005}  \at{Clustering by mixing flows}.
  \jt{Phys. Rev. Lett.}  \bvol{95},  \pg{240602}.

\bibitem[Einarsson {\em et~al.\/}(2014)Einarsson, Angilella \&
  Mehlig]{einarsson2014}
{\sc \au{Einarsson, J.}, \au{Angilella, J.~R.} \& \au{Mehlig, B.}} \yr{2014}
  \at{Orientational dynamics of weakly inertial axisymmetric particles in
  steady viscous flows}.  \jt{Physica D: Nonlinear Phenomena}  \bvol{278--279},
   \pg{79--85}.

\bibitem[Einarsson {\em et~al.\/}(2015)Einarsson, Candelier, Lundell, Angilella
  \& Mehlig]{einarsson2015b}
{\sc \au{Einarsson, J.}, \au{Candelier, F.}, \au{Lundell, F.}, \au{Angilella,
  J.R.} \& \au{Mehlig, B.}} \yr{2015}  \at{Effect of weak fluid inertia upon
  {Jeffery} orbits}.  \jt{Phys. Rev. E}  \bvol{91},  \pg{041002(R)}.

\bibitem[Einarsson {\em et~al.\/}(2016)Einarsson, Mihiretie, Laas, Ankardal,
  Angilella, Hanstorp \& Mehlig]{einarsson2016a}
{\sc \au{Einarsson, J.}, \au{Mihiretie, B.~M.}, \au{Laas, A.}, \au{Ankardal,
  S.}, \au{Angilella, J.~R.}, \au{Hanstorp, D.} \& \au{Mehlig, B.}} \yr{2016}
  \at{Tumbling of asymmetric microrods in a microchannel flow}.  \jt{Phys.
  Fluids}  \bvol{28},  \pg{013302}.

\bibitem[Fries {\em et~al.\/}(2017)Fries, Einarsson \& Mehlig]{Fries17}
{\sc \au{Fries, Johan}, \au{Einarsson, Jonas} \& \au{Mehlig, Bernhard}}
  \yr{2017}  \at{Angular dynamics of small crystals in viscous flow}.
  \jt{Phys. Rev. Fluids}  \bvol{2},  \pg{014302}.

\bibitem[Fries {\em et~al.\/}(2018)Fries, Kumar, Mihiretie, Hanstorp \&
  Mehlig]{Fries18}
{\sc \au{Fries, J.}, \au{Kumar, M.~Vijay}, \au{Mihiretie, B.~Mekonnen},
  \au{Hanstorp, D.} \& \au{Mehlig, B.}} \yr{2018}  \at{Spinning and tumbling of
  micron-sized triangles in a micro-channel shear flow}.  \jt{Phys. Fluids}
  \bvol{30},  \pg{033304}.

\bibitem[Guala {\em et~al.\/}(2005)Guala, L\"{u}thi, Liberzon, Tsinober \&
  Kinzelbach]{Guala}
{\sc \au{Guala, M.}, \au{L\"{u}thi, B.}, \au{Liberzon, A.}, \au{Tsinober, A.}
  \& \au{Kinzelbach, W.}} \yr{2005}  \at{On the evolution of material lines and
  vorticity in homogeneous turbulence}.  \jt{J. Fluid Mech.}  \bvol{533},
  \pg{339–359}.

\bibitem[Gustavsson {\em et~al.\/}(2016)Gustavsson, Berglund, Jonsson \&
  Mehlig]{Gus16}
{\sc \au{Gustavsson, K.}, \au{Berglund, F.}, \au{Jonsson, P.R.} \& \au{Mehlig,
  B.}} \yr{2016}  \at{Preferential sampling and small-scale clustering of
  gyrotactic microswimmers in turbulence}.  \jt{Phys. Rev. Lett.}  \bvol{116},
  \pg{108104}.

\bibitem[Gustavsson {\em et~al.\/}(2014)Gustavsson, Einarsson \& Mehlig]{Gus14}
{\sc \au{Gustavsson, K.}, \au{Einarsson, J.} \& \au{Mehlig, B.}} \yr{2014}
  \at{Tumbling of small axisymmetric particles in random and turbulent flows}.
  \jt{Phys. Rev. Lett.}  \bvol{112},  \pg{014501}.

\bibitem[Gustavsson \& Mehlig(2016)]{Gus16:rev}
{\sc \au{Gustavsson, K.} \& \au{Mehlig, B.}} \yr{2016}  \at{Statistical models
  for spatial patterns of heavy particles in turbulence}.  \jt{Advances in
  Physics}  \bvol{65},  \pg{1--57}.

\bibitem[Hejazi {\em et~al.\/}(2017)Hejazi, Mehlig \& Voth]{Hejazi2017}
{\sc \au{Hejazi, Bardia}, \au{Mehlig, Bernhard} \& \au{Voth, Greg~A.}}
  \yr{2017}  \at{Emergent scar lines in chaotic advection of passive
  directors}.  \jt{Phys. Rev. Fluids}  \bvol{2},  \pg{124501}.

\bibitem[Hinch \& Leal(1972)]{Hinch1972}
{\sc \au{Hinch, E.~J.} \& \au{Leal, L.~G.}} \yr{1972}  \at{The effect of
  {B}rownian motion on the rheological properties of a suspension of
  non-spherical particles}.  \jt{J. Fluid Mech.}  \bvol{52},  \pg{683--712}.

\bibitem[Jeffery(1922)]{Jeffery:1922}
{\sc \au{Jeffery, G.B.}} \yr{1922}  \at{The motion of ellipsoidal particles
  immersed in a viscous fluid}.  \jt{Proceedings of the Royal Society of London
  A: Mathematical, Physical and Engineering Sciences}  \bvol{102},
  \pg{161--179}.

\bibitem[Kim {\em et~al.\/}(1987)Kim, Moin \& Moser]{Kim87}
{\sc \au{Kim, John}, \au{Moin, Parviz} \& \au{Moser, Robert}} \yr{1987}
  \at{Turbulence statistics in fully developed channel flow at low {Reynolds}
  number}.  \jt{J. Fluid Mech.}  \bvol{177},  \pg{133--166}.

\bibitem[Lundell {\em et~al.\/}(2011)Lundell, S{\"o{}}derberg \&
  Alfredsson]{Lun11}
{\sc \au{Lundell, F.}, \au{S{\"o{}}derberg, D.} \& \au{Alfredsson, H.}}
  \yr{2011}  \at{Fluid mechanics of papermaking}.  \jt{Annu.~Rev.~Fluid~Mech.}
  \bvol{43},  \pg{195--217}.

\bibitem[Mansour {\em et~al.\/}(1988)Mansour, Kim \& Moin]{Man88}
{\sc \au{Mansour, N.~N.}, \au{Kim, J.} \& \au{Moin, P.}} \yr{1988}
  \at{Reynolds-stress and dissipation-rate budgets in a turbulent channel
  flow}.  \jt{Journal of Fluid Mechanics}  \bvol{194},  \pg{15–44}.

\bibitem[Marchioli {\em et~al.\/}(2010)Marchioli, Fantoni \& Soldati]{Mar10}
{\sc \au{Marchioli, Cristian}, \au{Fantoni, Marco} \& \au{Soldati, Alfredo}}
  \yr{2010}  \at{Orientation, distribution, and deposition of elongated,
  inertial fibers in turbulent channel flow}.  \jt{Phys. Fluids}  \bvol{22},
  \pg{033301}.

\bibitem[Marchioli \& Soldati(2002)]{Mar02}
{\sc \au{Marchioli, Cristian} \& \au{Soldati, Alfredo}} \yr{2002}
  \at{Mechanisms for particle transfer and segregation in a turbulent boundary
  layer}.  \jt{J. Fluid Mech.}  \bvol{468},  \pg{283--315}.

\bibitem[Meibohm {\em et~al.\/}(2017)Meibohm, Pistone, Gustavsson \&
  Mehlig]{Mei17}
{\sc \au{Meibohm, J.}, \au{Pistone, L.}, \au{Gustavsson, K.} \& \au{Mehlig,
  B.}} \yr{2017}  \at{Relative velocities in bidisperse turbulent suspensions}.
   \jt{Phys. Rev. E}  \bvol{96},  \pg{061102}.

\bibitem[Mortensen {\em et~al.\/}(2008)Mortensen, Andersson, Gillissen \&
  Boersma]{Mor08}
{\sc \au{Mortensen, P.~H.}, \au{Andersson, H.~I.}, \au{Gillissen, J. J.~J.} \&
  \au{Boersma, B.~J.}} \yr{2008}  \at{Dynamics of prolate ellipsoidal particles
  in a turbulent channel flow}.  \jt{Phys. Fluids}  \bvol{20},  \pg{093302}.

\bibitem[Ni {\em et~al.\/}(2014)Ni, Ouelette \& Voth]{Ni14}
{\sc \au{Ni, R.}, \au{Ouelette, N.~T.} \& \au{Voth, G.~A.}} \yr{2014}
  \at{Alignment of vorticity and rods with {L}agrangian fluid stretching in
  turbulence}.  \jt{J. Fluid Mech.}  \bvol{743},  \pg{R3}.

\bibitem[Parsa {\em et~al.\/}(2012)Parsa, Calzavarini, Toschi \& Voth]{Par12}
{\sc \au{Parsa, S.}, \au{Calzavarini, E.}, \au{Toschi, F.} \& \au{Voth, G.~A.}}
  \yr{2012}  \at{Rotation rate of rods in turbulent fluid flow}.  \jt{Phys.
  Rev. Lett.}  \bvol{109},  \pg{134501}.

\bibitem[Parsa {\em et~al.\/}(2011)Parsa, Guasto, Kishore, Ouellette, Gollub \&
  Voth]{Par11}
{\sc \au{Parsa, Shima}, \au{Guasto, Jeffrey~S.}, \au{Kishore, Monica},
  \au{Ouellette, Nicholas~T.}, \au{Gollub, J.~P.} \& \au{Voth, Greg~A.}}
  \yr{2011}  \at{Rotation and alignment of rods in two-dimensional chaotic
  flow}.  \jt{Phys. Fluids}  \bvol{23},  \pg{043302}.

\bibitem[Pumir(2017)]{Pum17}
{\sc \au{Pumir, Alain}} \yr{2017}  \at{Structure of the velocity gradient
  tensor in turbulent shear flows}.  \jt{Phys. Rev. Fluids}  \bvol{2},
  \pg{074602}.

\bibitem[Pumir \& Wilkinson(2011)]{Pum11}
{\sc \au{Pumir, A.} \& \au{Wilkinson, M.}} \yr{2011}  \at{Orientation
  statistics of small particles in turbulence}.  \jt{New Journal of Physics}
  \bvol{13},  \pg{093030}.

\bibitem[Ros\'{e}n {\em et~al.\/}(2015)Ros\'{e}n, Einarsson, Nordmark, Aidun,
  Lundell \& Mehlig]{rosen2015d}
{\sc \au{Ros\'{e}n, T.}, \au{Einarsson, J.}, \au{Nordmark, A.}, \au{Aidun,
  C.~K.}, \au{Lundell, F.} \& \au{Mehlig, B.}} \yr{2015}  \at{Numerical
  analysis of the angular motion of a neutrally buoyant spheroid in shear flow
  at small {R}eynolds numbers}.  \jt{Phys. Rev. E}  \bvol{92}, 063022.

\bibitem[Subramanian \& Koch(2005)]{subramanian2005}
{\sc \au{Subramanian, G.} \& \au{Koch, D.~L.}} \yr{2005}  \at{Inertial effects
  on fibre motion in simple shear flow}.  \jt{J. Fluid Mech.}  \bvol{535},
  \pg{383--414}.

\bibitem[Turitsyn(2007)]{Tur07}
{\sc \au{Turitsyn, K.~S.}} \yr{2007}  \at{Polymer dynamics in chaotic flows
  with a strong shear component}.  \jt{Journal of Experimental and Theoretical
  Physics}  \bvol{105},  \pg{655--664}.

\bibitem[Voth \& Soldati(2017)]{Voth16}
{\sc \au{Voth, G.} \& \au{Soldati, A.}} \yr{2017}  \at{Anisotropic particles in
  turbulence}.  \jt{Annu. Rev. Fluid Mech.}  \bvol{49},  \pg{249--276}.

\bibitem[Wilkinson {\em et~al.\/}(2009)Wilkinson, Bezuglyy \& Mehlig]{Wil09}
{\sc \au{Wilkinson, M.}, \au{Bezuglyy, V.} \& \au{Mehlig, B.}} \yr{2009}
  \at{Fingerprints of random flows?}  \jt{Phys. Fluids}  \bvol{21},
  \pg{043304}.

\bibitem[Wilkinson {\em et~al.\/}(2011)Wilkinson, Bezuglyy \& Mehlig]{Wil11}
{\sc \au{Wilkinson, M.}, \au{Bezuglyy, V.} \& \au{Mehlig, B.}} \yr{2011}
  \at{Emergent order in rheoscopic swirls}.  \jt{J. Fluid Mech.}  \bvol{667},
  \pg{158}.

\bibitem[Xu {\em et~al.\/}(2011)Xu, Pumir \& Bodenschatz]{xu2011}
{\sc \au{Xu, H.}, \au{Pumir, A.} \& \au{Bodenschatz, E.}} \yr{2011}  \at{The
  pirouette effect in turbulent flows}.  \jt{Nature Physics}  \bvol{7},
  \pg{709}.

\bibitem[Zhao \& Andersson(2016)]{Zha16}
{\sc \au{Zhao, Lihao} \& \au{Andersson, Helge~I.}} \yr{2016}  \at{Why spheroids
  orient preferentially in near-wall turbulence}.  \jt{J. Fluid Mech.}
  \bvol{807},  \pg{221--234}.

\bibitem[Zhao {\em et~al.\/}(2015)Zhao, Challabotla, Andersson \&
  Variano]{Zha15}
{\sc \au{Zhao, L.}, \au{Challabotla, N.~R.}, \au{Andersson, H.~I.} \&
  \au{Variano, E.~A.}} \yr{2015}  \at{Rotation of nonspherical particles in
  turbulent channel flow}.  \jt{Phys. Rev. Lett.}  \bvol{115},  \pg{244501}.

\bibitem[Zhao {\em et~al.\/}(2019)Zhao, Gustavsson, Ni, Kramel, Voth, Andersson
  \& Mehlig]{Zha19}
{\sc \au{Zhao, L.}, \au{Gustavsson, K.}, \au{Ni, R.}, \au{Kramel, S.},
  \au{Voth, G.}, \au{Andersson, H.~I.} \& \au{Mehlig, B.}} \yr{2019}
  \at{Passive directors in turbulence}.  \jt{Phys. Rev. Fluids}  \bvol{4},
  \pg{054602}.

\end{thebibliography}
\end{document}